\documentclass[a4paper,11pt]{article}
\pdfoutput=1
\usepackage{jheppub}

\usepackage{epsf,epsfig,subfigure,graphicx,amsmath,amssymb,mathtools,hhline}
\usepackage{color}
\usepackage{float}
\usepackage{multirow}
\usepackage{nicefrac}
\usepackage{epstopdf,slashed,xcolor}

\usepackage{soul}

\newcommand{\be}{\begin{eqnarray}}
\newcommand{\ee}{\end{eqnarray}}

\newcommand{\ba}{\begin{array}}
\newcommand{\ea}{\end{array}}
\newcommand{\bee}{\begin{equation}\ba{c}}
\newcommand{\eee}{\ea\end{equation}}

\newcommand{\bi}{\begin{itemize}}
\newcommand{\ei}{\end{itemize}}

\toccontinuoustrue

\title{ 
New constraints and discovery potential for Higgs to Higgs cascade decays through vectorlike leptons 
}
\author{Radovan Derm\'i\v{s}ek$^{1,2}$,}
\author{Enrico Lunghi$^1$}
\author{and Seodong Shin$^{1,3}$}
\affiliation{
$^1$Physics Department, Indiana University, Bloomington, IN 47405, USA \\
$^2$Department of Physics and Astronomy and Center for Theoretical Physics, Seoul National University, Seoul 151-747, Korea \\
$^3$Department of Physics \& IPAP, Yonsei University, Seoul 03722, Korea \\
}
\emailAdd{dermisek@indiana.edu} 
\emailAdd{elunghi@indiana.edu} 
\emailAdd{shinseod@indiana.edu}

\abstract{
One of the cleanest signatures of a heavy Higgs boson in models with vectorlike leptons is $H\to e_4^\pm \ell^\mp \to h\ell^+\ell^-$ which, in two Higgs doublet model type-II, can  even be the dominant decay mode of heavy Higgses. 
 Among the decay modes of the standard model like Higgs boson, $h$, we consider $b \bar b$ and $\gamma \gamma$ as representative channels with sizable and negligible background, respectively. We obtained new model independent limits on production cross section for this process from recasting existing experimental searches and interpret them within the two Higgs doublet model. In addition, we show that these limits can be improved by about two orders of magnitude with appropriate selection cuts immediately with existing data sets. We also discuss expected sensitivities with  integrated luminosity up to 3 ab$^{-1}$ and present a brief overview of other channels. 
\newpage
}

\preprint{
\begin{minipage}{3cm}
\small
\flushright
IU-HET-614
\end{minipage}} 

\begin{document}

\maketitle

\section{Introduction}
\label{sec:intro}
In models with vectorlike fermions, even a very small mixing with one of the Standard Model (SM)  families forces  the lightest vectorlike eigenstate to decay into $W/Z/h$ and a SM fermion.  If there are more Higgs bosons, as in models with extended Higgs sector, the same mixing allows the heavy Higgses to decay into a vectorlike and a SM fermion. This leads to many new opportunities to search for new Higgs bosons and vectorlike matter simultaneously~\cite{Dermisek:2015hue}. 

Limits from direct searches for vectorlike leptons are significantly weaker than for vectorlike quarks~\cite{Dermisek:2015hue,Dermisek:2015oja,Dermisek:2014qca,Kumar:2015tna,Falkowski:2013jya,Ellis:2014dza}. In addition, leptons in final states typically result in clean signatures. Thus searching for combined signatures of vectorlike leptons and new Higgs bosons is especially advantageous. 
In this work, we focus on the process:
\begin{align}
pp \to H \to e_4^\pm \mu^\mp \to h \mu^+ \mu^- \; ,  
\label{eq:process}
\end{align}
where $H$ is the heavy CP even Higgs and the $e_4$ is a new charged lepton (note that, in a small region of the parameter space $e_4^\pm \mu^\mp$ is also a possible decay mode for the  SM Higgs~\cite{Dermisek:2014cia}). We obtain new constraints on this process from recasting existing experimental searches and find future experimental sensitivities by optimizing the selection cuts. 

This process appears for example in a two Higgs doublet model type-II with vectorlike leptons mixing with second SM family introduced in ref.~\cite{Dermisek:2015oja, {Dermisek:2013gta}} and it was identified as one of the cleanest signatures of heavy Higgses in this class of models~\cite{Dermisek:2015hue}. It was found that $H \to e_4^\pm \mu^\mp$ can be the dominant decay mode of the heavy Higgs in a large range of parameters~\cite{Dermisek:2015hue}. Moreover, as we will show, the high luminosity LHC is sensitive to this process even for branching ratio $\sim 10^{-5}$.\footnote{Considering mixing with the first SM family, or considering the CP odd Higgs instead of $H$ would lead to very similar results. }

\begin{table}
\begin{center}
\begin{tabular}{|l|l|c|}
\hline
Higgs decay mode & final state & $\sigma$ \cr \hline
$h\to b \bar b$ & $b \bar b \mu^+ \mu^-$ & 5.3 pb \cr
$h\to \tau^+ \tau^-$ & $\tau^+ \tau^- \mu^+ \mu^-$ & 0.58 pb \cr
$h\to WW^* \to \ell^+ \ell^- \nu_\ell \bar{\nu}_\ell$ ($\ell = e,\; \mu$) & $\ell^+ \ell^- \mu^+ \mu^- \nu_\ell \bar{\nu}_\ell$ & 97 fb \cr
$h\to \gamma \gamma$ & $\gamma\gamma\mu^+\mu^-$ & 28 fb \cr
$h\to \mu^+\mu^-$ & $\mu^+\mu^-\mu^+\mu^-$ & 2 fb \cr
$h\to ZZ^* \to 2\ell^+ 2\ell^-$ ($\ell = e,\; \mu$) & $\ell^+\ell^-\ell^+\ell^-\mu^+\mu^-$ & 1.1 fb \cr
\hline
\end{tabular}
\caption{The 13 TeV LHC production rates for $H \to h \mu^+ \mu^-$ for various decay channels of the SM Higgs boson in two Higgs doublet model type-II  for $m_H = 200$ GeV, $\tan\beta=1$ and ${\rm BR} (H \to e_4^\pm \mu^\mp \to h\mu^+\mu^-) = 0.5$. The value for $h\to \mu^+\mu^-$ assumes that the $\mu-\mu-h$ Yukawa coupling is not modified; in our model however it can be suppressed or enhanced, see ref.~\cite{Dermisek:2013gta}. }
\label{table:channel}
\end{center}
\end{table}

Depending on the decay mode of the SM-like  Higgs boson, $h$, the process (\ref{eq:process}) leads to several interesting final states with rates summarized in table~\ref{table:channel} for a  representative set of parameters: $m_H = 200$ GeV, $\tan\beta=1$, ${\rm BR} (H \to e_4^\pm \mu^\mp \to h\mu^+ \mu^-) = 0.5$.
Each decay mode of the SM-like Higgs boson $h \to b \bar b, WW^\ast, ZZ^\ast, \gamma \gamma, \tau^+ \tau^-, \mu^+ \mu^-$ provides its unique signal~\cite{Dermisek:2015hue}. 
A prominent feature of all these channels is that the dimuon pair produced with the SM Higgs does not peak at the $Z$ boson invariant mass as is the case for most backgrounds. Moreover, in most channels, it is possible to reconstruct the $H$ and $e_4$ masses.

Although specific searches for the process (\ref{eq:process}) do not exist, the particle content in final states is the same as for  $pp \to Zh$  or $pp \to A \to Zh$ and thus related Higgs searches constrain our process. We recast  experimental searches  for $A \to hZ \to b \bar b \ell^+ \ell^-$, where $A$ is a heavy new particle and $\ell = e, \mu$, performed at ATLAS~\cite{Aad:2015yza} and CMS~\cite{Khachatryan:2015lba} and for $p p \to h \ell X \to \gamma \gamma \ell X$~\cite{Aad:2014lwa} and $pp \to Z \gamma \gamma \to \ell^+ \ell^- \gamma \gamma$~\cite{Aad:2016sau} performed at ATLAS. We set model independent limits on production cross section of (\ref{eq:process}) in $b\bar b\mu^+\mu^-$ and $\gamma\gamma\mu^+\mu^-$ final states as functions of masses of $H$  and $e_4$. Then we suggest a simple modification of existing searches, the addition of the ``off-$Z$" cut, which takes advantage of the two muons in the final state not originating from a $Z$ boson, and show how the limits could be improved immediately with current data and indicate experimental sensitivities with future data sets.

After deriving model independent limits we interpret them within the two Higgs doublet model type-II. We first use the scan of the parameter space of this model for $m_H <2m_t$ presented  in ref.~\cite{Dermisek:2015hue}, where constraints from electroweak precision observables (oblique corrections, muon lifetime, $Z$-pole observables, $W\to\mu\nu$), constraints on pair production of vectorlike leptons obtained from searches for anomalous production of multilepton events~\cite{Dermisek:2014qca}, $H\to (WW,\gamma\gamma)$ and $h \to \gamma \gamma$~\cite{Dermisek:2015vra, Dermisek:2015hue} have been included. In addition,  we extend the scan for $m_H > 2m_t$ and  whole range of $\tan \beta$. We show how current experimental studies constrain the allowed parameter space and what we can achieve by means of optimized search strategies.  

This paper is organized as follows. In section~\ref{sec:anal} we briefly summarize our analysis method such as implementing the event simulations and setting the limits on our parameter space. The new constraints recasted from the existing searches are shown in section~\ref{sec:8data} and the expected experimental sensitivities in the future with our suggested cuts are discussed in section~\ref{sec:sensitivity}. We study the impact of the new constraints and future prospects of existing and suggested searches on the two Higgs doublet model type-II with vectorlike leptons in section~\ref{sec:results}. We further analyze the parameters of heavy Higgs above the $t \bar t$ threshold in section~\ref{sec:heavyH}. Finally we give conclusions in section~\ref{sec:conclusions}.

\section{Analysis method}
\label{sec:anal}
In this section we discuss the tools used for the event simulation and the statistical approach we adopt to set the limits. The new physics model is implemented in {\tt FeynRules}~\cite{Degrande:2011ua}, events are generated with {\tt MadGraph5}~\cite{Alwall:2014hca} and showered with {\tt Pythia6}~\cite{Sjostrand:2006za}. The resulting {\tt StdHEP} event files are converted into CERN {\tt root} format using {\tt Delphes}~\cite{deFavereau:2013fsa}. Jets are identified using the anti-$k_t$ algorithm of {\tt FastJet}~\cite{Cacciari:2005hq,Cacciari:2011ma} with angular separation $\Delta R = 0.4$.

We present 95\% C.L. upper limits calculated   using a modified frequentist construction (CL$_{\rm s}$)~\cite{cls1, cls2}. In recasting the searches presented in refs.~\cite{Aad:2015yza,Aad:2014lwa,Aad:2016sau}, we follow the method described in refs.~\cite{Dermisek:2014qca,Dermisek:2013cxa} where the Poisson likelihood is assumed. In order to calculate the number of events, $N_s^{95}$, that corresponds to the 95\% C.L. upper limits, we consider a set of event numbers ($\{ n_i \}$) corresponding to a Poisson distribution with expectation value $b$ equal to the number of background events. For each $n_i$ the signal-plus-background hypothesis is tested using the CL$_{\rm s}$ method. The expected upper limit $N_s^{95}$ is the median of the $\{ n_i \}$ that pass the test.\footnote{The median discovery significance is obtained by testing the background-only hypothesis with a data set with expectation value $s + b$ where $s$ is the number of expected signal events.} 

The 95\% C.L. upper limits on the total $pp\to H\to h\mu^+\mu^-$ cross section normalized to the production cross section of a SM-like heavy Higgs ($H_{\rm SM}$) are given by 
\begin{align}
\frac{\sigma(p p \to H)}{\sigma(p p \to H_{\rm SM})} {\rm BR}(H \to h \mu^+ \mu^-) < 
\begin{cases}
\frac{N_s^{95} (bb)}{\mathcal L \cdot \xi_{bb} \cdot A_{{\rm NP} bb}} \cdot \frac{1}{\sigma(p p \to H_{\rm SM}) \times {\rm BR}(h \to b \bar b)}  \cr
\frac{N_s^{95} (\gamma \gamma)}{\mathcal L \cdot \xi_{\gamma \gamma} \cdot A_{{\rm NP} \gamma \gamma}} \cdot \frac{1}{\sigma(p p \to H_{\rm SM}) \times {\rm BR}(h \to \gamma \gamma)}
\end{cases},
\label{eq:sensitivity}
\end{align}
where $A_{{\rm NP} bb}$ and $A_{{\rm NP} \gamma \gamma}$ are the MC level acceptances (calculated using the selection cuts of the analyses that we recast) of the $b \bar b \mu^+ \mu^-$ and $\gamma \gamma \mu^+ \mu^-$ channels, $\xi_{bb}$ and $\xi_{\gamma \gamma}$ are the detector level efficiencies and $\mathcal L$ is the integrated luminosity. 

Note that the negligible background to the $\gamma \gamma \mu^+ \mu^-$ mode implies $N_s^{95} (\gamma \gamma) = 3$ (with Poisson statistics, a null observation over a null background is compatible with up to three signal events at 95\% C.L.~\cite{Dermisek:2014qca, Dermisek:2013cxa}).\footnote{With 3 ab$^{-1}$ at 13 TeV the number of background events is non-zero and we take this into account in our limit setting.} For this reason the $b \bar b \mu^+ \mu^-$ can provide a stronger constraint as long as 
\begin{align}
N_s^{95} (bb) < N_s^{95}(\gamma\gamma)\frac{{\rm BR}(h \to b \bar b) \; \xi_{bb} \; A_{{\rm NP} bb}}{{\rm BR}(h \to \gamma \gamma) \; \xi_{\gamma \gamma} \;A_{{\rm NP} \gamma \gamma}} \sim 759 \times \frac{\xi_{bb} \; A_{{\rm NP} bb}}{\xi_{\gamma \gamma} \; A_{{\rm NP} \gamma \gamma}}~,
\end{align}
where the ratio of experimental efficiencies ($\xi_{bb}/\xi_{\gamma\gamma}$) is about one, the ratio of Monte Carlo level acceptances ($A_{{\rm NP} bb}/A_{{\rm NP} \gamma\gamma}$) varies between one and three, $N_s^{95}(\gamma\gamma)$ is almost constant, and $N_s^{95}(bb)$ increases with the integrated luminosity as can be seen in table~\ref{table:newbb}.

\section{New constraints from the 8 TeV LHC data}
\label{sec:8data}
In this section we extract upper bounds on the heavy Higgs cascade decays we consider from existing searches with 20.3 fb$^{-1}$ of integrated luminosity at 8 TeV. The process $H \to h \mu^+ \mu^- \to b \bar b \mu^+ \mu^-$  is constrained by searches for $A \to hZ \to b \bar b \ell^+ \ell^-$, where $A$ is a heavy new particle and $\ell = e, \mu$. These searches have been performed at ATLAS~\cite{Aad:2015yza}\footnote{See also ref.~\cite{13hz} for resonances heavier than $\gtrsim 500$ GeV.} and CMS~\cite{Khachatryan:2015lba} (we focus on the former because they provide the explicit number of observed and expected events, allowing us to investigate the impact of the different cuts). The process $H \to h \mu^+ \mu^- \to \gamma \gamma \mu^+ \mu^-$ is constrained by the $h \to \gamma \gamma$ ATLAS search~\cite{Aad:2014lwa} where the results with an  inclusive lepton cut are presented ($p p \to h \ell X \to \gamma \gamma \ell X$) and also $pp \to Z \gamma \gamma \to \ell^+ \ell^- \gamma \gamma$~\cite{Aad:2016sau}. 

The results that we obtain and describe in details in the next three subsections are presented in figures~\ref{fig:bbmm8bound}-\ref{fig:dpmm8bisbound}. The constraints on $\left(\sigma_H / \sigma_{H_{\rm SM}}\right) \times {\rm BR}(H \to h \mu^+ \mu^-)$ are mostly constant as a function of the $H$ and $e_4$ masses and vary in the range $[0.1,0.3]$. We steeply loose sensitivity for $e_4$ close in mass to either the SM or the heavy Higgs (the transverse momentum of one the muons becomes too soft), or for a lighter heavy Higgs (the maximum value of the dilepton invariant mass is $m_H-m_h$, see eq.~(\ref{dilepton}) and the related discussion, and the requirement of an on-shell $Z$ cuts all signal events for small $m_H$).

\subsection{Recast of the $b \bar b \mu^+ \mu^-$ search}
\label{sec:bbmm8}
From the results presented in ref.~\cite{Aad:2015yza} we extract the observed upper limit $N_s^{95} (bb)$. We extract the detector level efficiency $\xi_{bb}$ by comparing the expected number of the Higgsstrahlung ($p p \to hZ$) events given in ref.~\cite{Aad:2015yza} to the fiducial number of events that we calculate. In this way our $\xi_{bb}$ includes the effect of the profile likelihood fit of MC background events to the data in the control region. Using the Higgsstrahlung cross section presented in refs.~\cite{Heinemeyer:2013tqa,Agashe:2014kda} and the acceptances we calculate, we find $\xi_{bb} \simeq 32$\%.

The fiducial region adopted in ref.~\cite{Aad:2015yza} is defined as follows. The two muons are required to have pseudorapidity $|\eta| < 2.5$ and transverse momenta larger than 25 and 7 GeV. Their invariant mass is required to lie in the range $83 \; {\rm GeV} < m_{\ell \ell} < 99 \; {\rm  GeV}$; note that this requirement cuts out a large part of our signal because we do not have an on-shell $Z$.  A missing transverse energy cut $E_T^{\rm miss} < 60$ GeV is imposed to reject the $t \bar t$ background. In order to reduce the $Z$+jets background the transverse momentum of the dilepton system ($p_T^Z$) is required to satisfy  $p_T^Z > 0.4 \times m_{VH} - 100\,{\rm GeV}$ where $m_{VH}$ is the invariant mass of the two leptons and two $b$-jets. The two $b$ tagged-jets are required to have $|\eta| < 2.5$ and $p_T > 45, 30$ GeV to suppress $Z$+jets background. The invariant mass of the $b \bar b$ system is required to lie in the range $105 \; {\rm GeV}< m_{b \bar b} < 145 \; {\rm GeV}$. Finally, in order to improve the resolution of $m_{VH}$, the Higgs boson candidate jet momenta are scaled by $m_h / m_{b \bar b}$ where $m_h = 125$ GeV.

\begin{figure}
\begin{center}
\includegraphics[width=.495\linewidth]{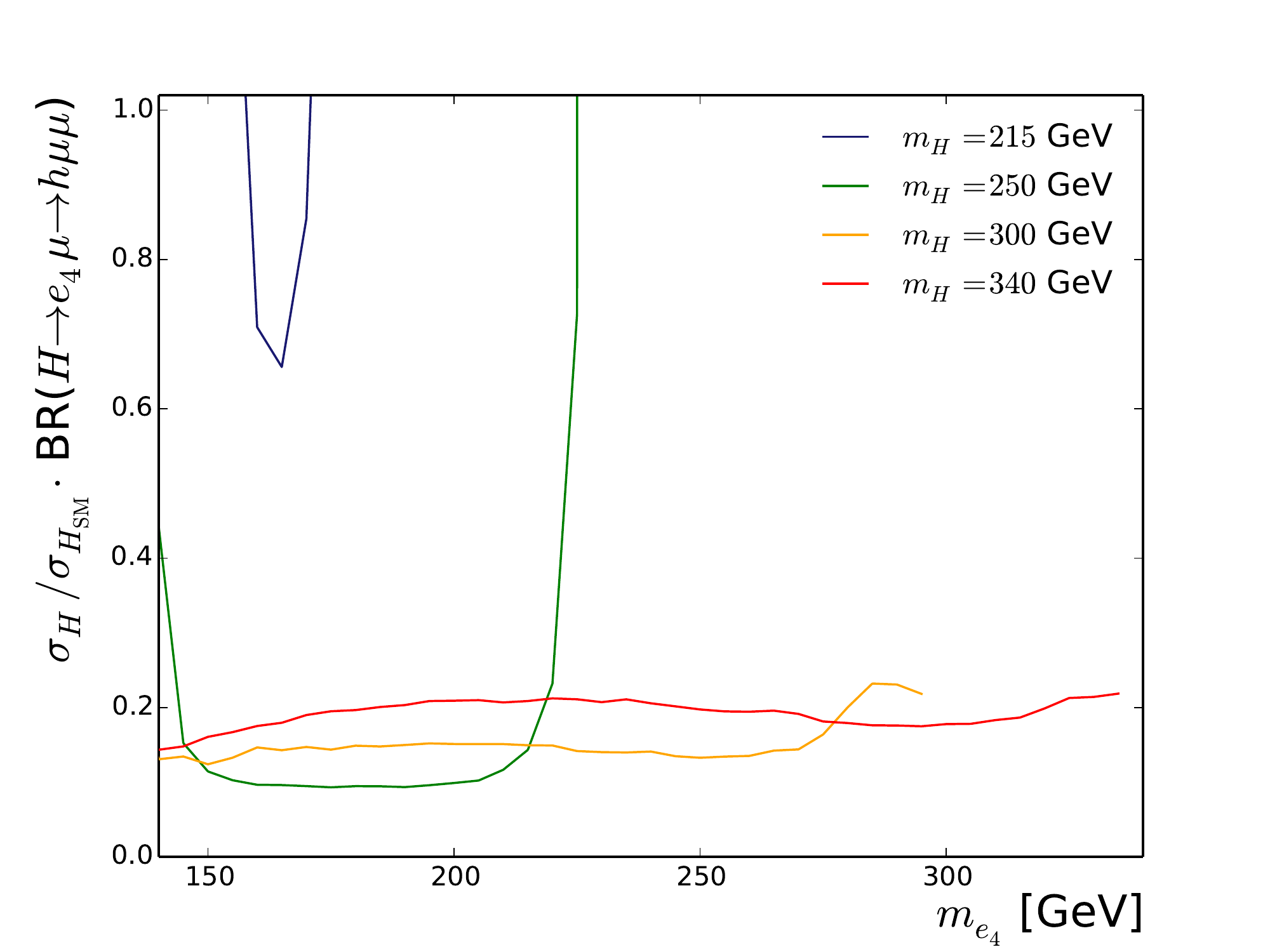}
\includegraphics[width=.495\linewidth]{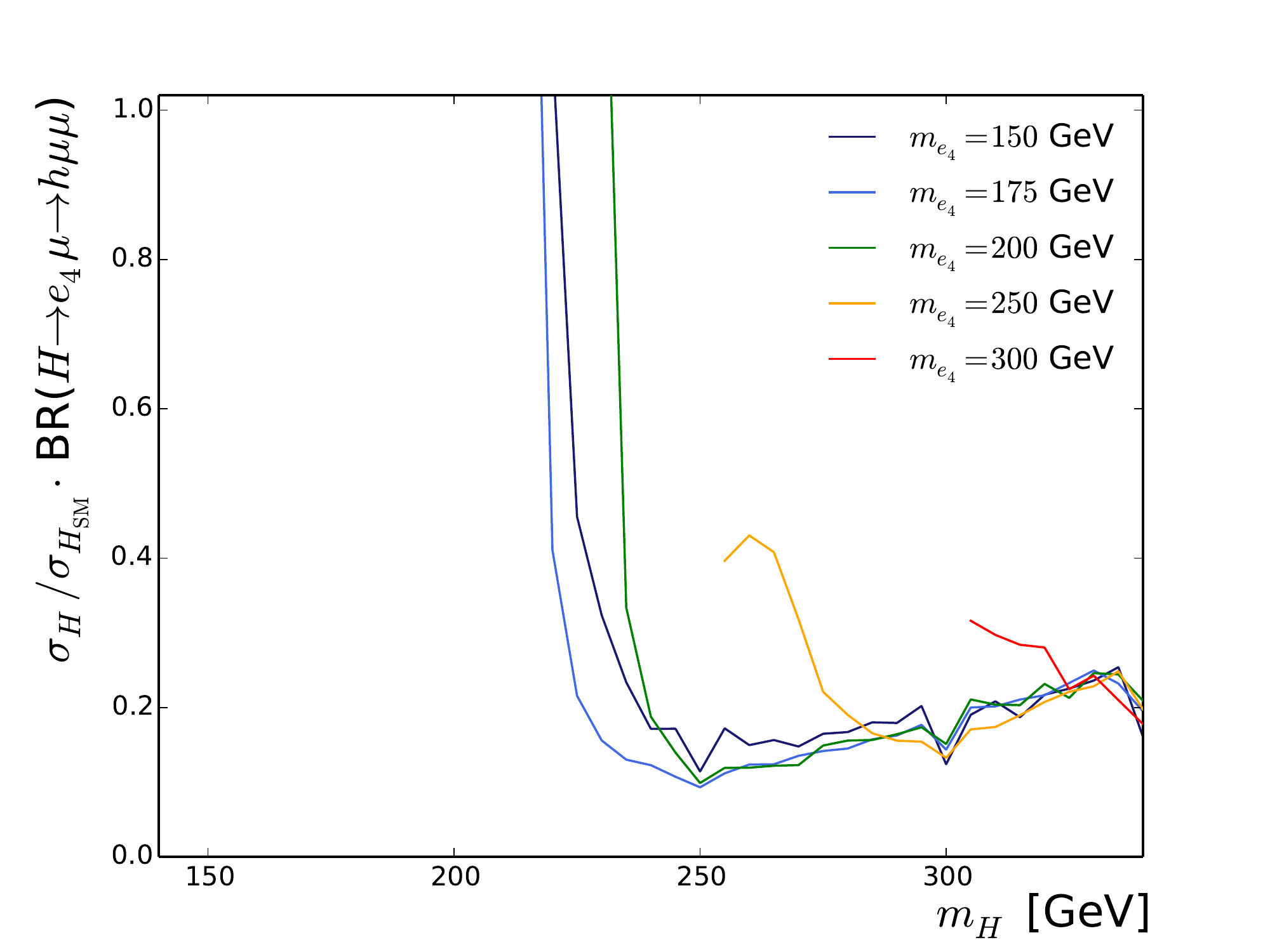}
\caption{ The 95\% CL upper bounds on $pp \to H \to e_4^\pm \mu^\mp \to h \mu^+ \mu^-$ for various choices of $m_H$ and $m_{e_4}$ obtained by recasting the ATLAS search for $A \to hZ \to b \bar b \ell^+ \ell^-$~\cite{Aad:2015yza} from the 8 TeV full data.
}
\label{fig:bbmm8bound}
\end{center}
\end{figure}

Using the observed and expected background events given in table~1 of ref.~\cite{Aad:2015yza} we obtain $N_s^{95} (b b) \simeq 88$. In figure~\ref{fig:bbmm8bound} we present the upper limits on $pp \to H \to e_4^\pm \mu^\mp \to h \mu^+ \mu^-$ for various choices of $m_H$ and $m_{e_4}$. The limits become very weak for $m_H \lesssim 215$ GeV because of the hard lepton selection cuts. Note that in the type-II two Higgs doublet model the ratio of Higgs production cross sections, that we show on the vertical axes, depends on $\tan\beta$. For $\tan\beta < 7$ this ratio is given by $\cot^2\beta$ to a good approximation. At larger values of $\tan\beta$ the bottom Yukawa coupling increases implying a non-negligible impact on the $b\bar b$ and gluon fusion production cross sections. We express our result in terms of $\left(\sigma_H / \sigma_{H_{\rm SM}}\right) \times {\rm BR}(H \to h \mu^+ \mu^-)$ because the limits on this quantity are model independent.

\subsection{Recast of the $\gamma \gamma \mu X$ search}
\label{sec:dpmm8}
%

\begin{figure}
\begin{center}
\includegraphics[width=.495\linewidth]{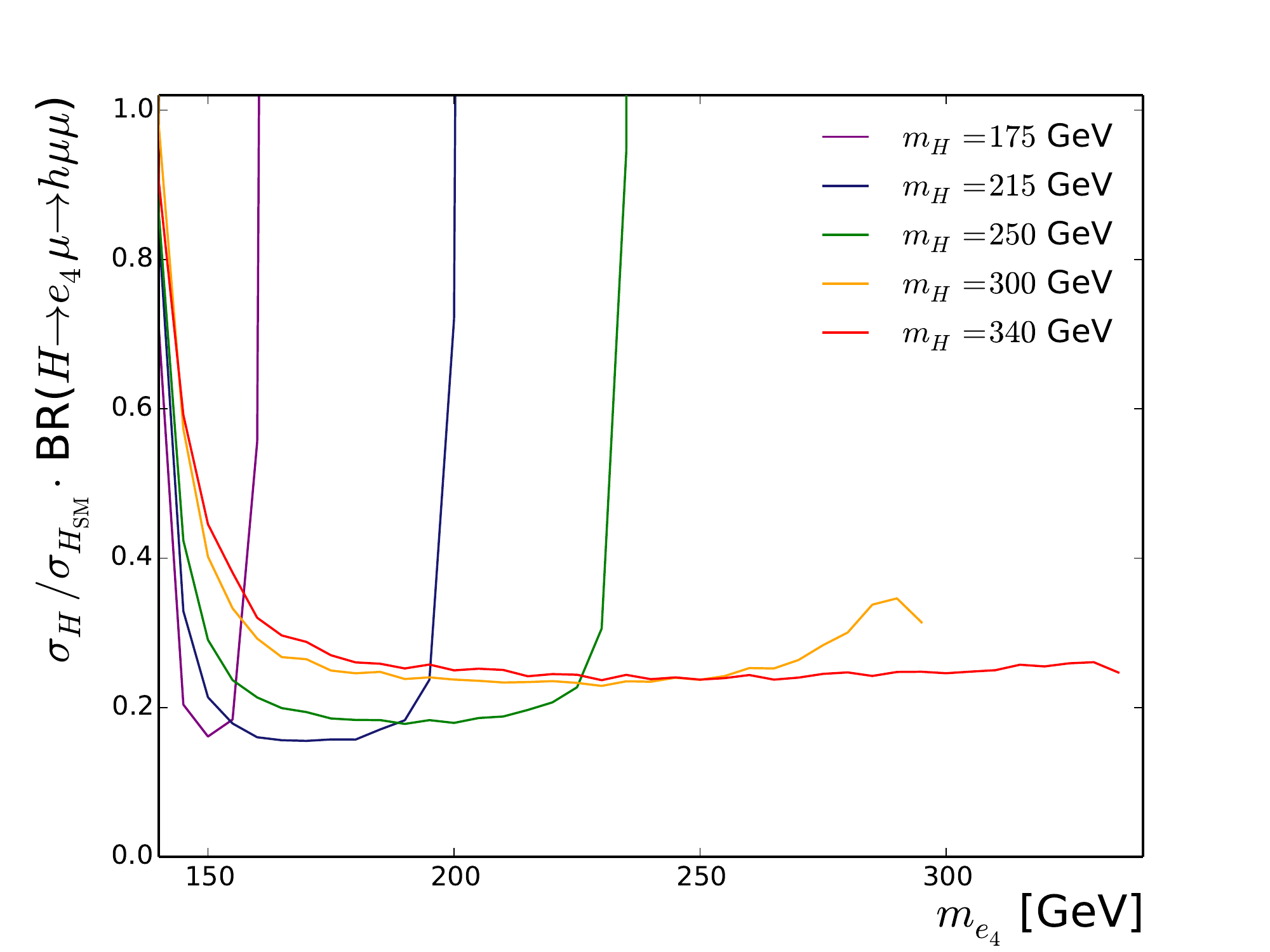}
\includegraphics[width=.495\linewidth]{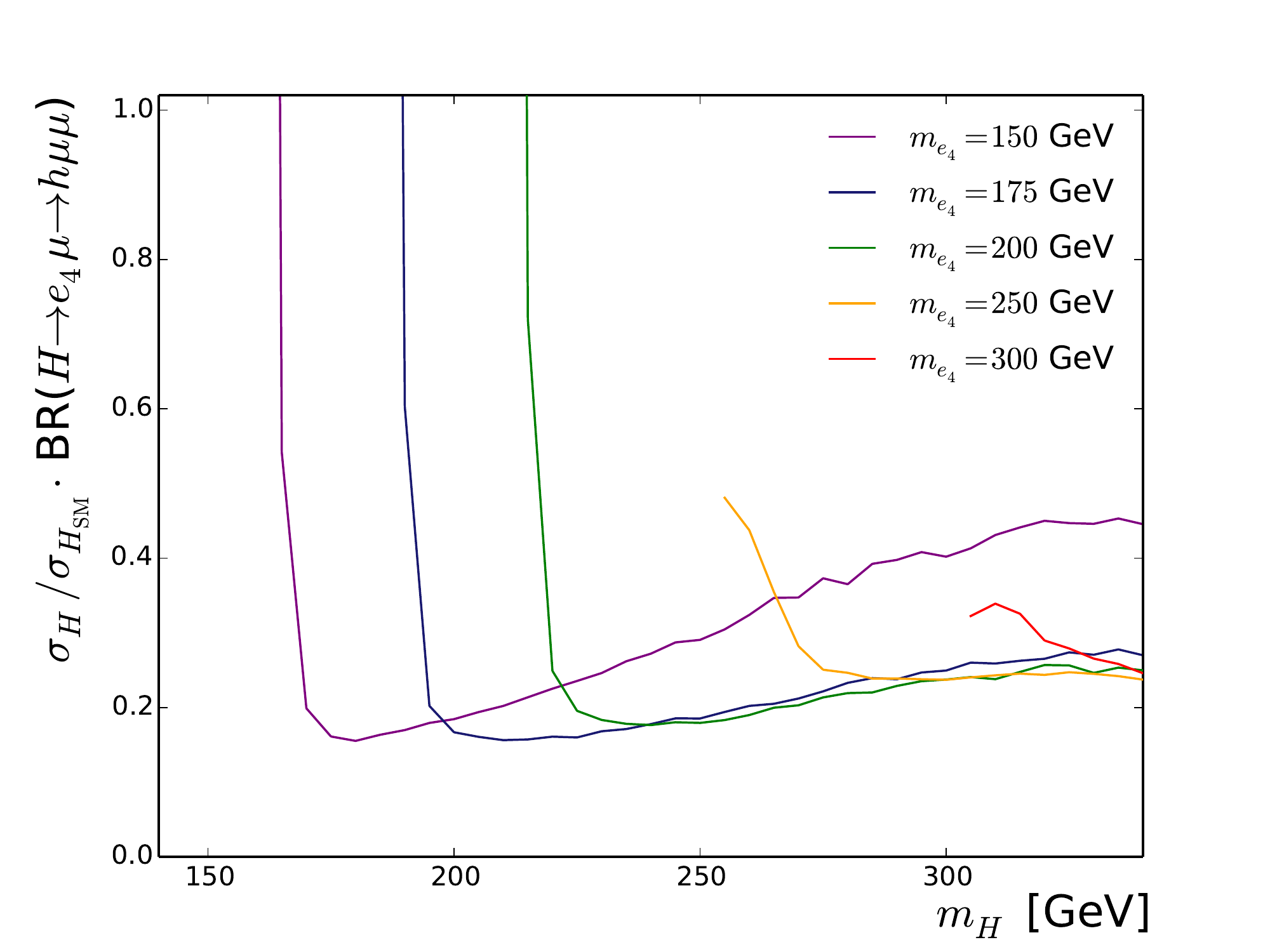}
\caption{The 95\% C.L. upper bounds on $pp \to H \to e_4^\pm \mu^\mp \to h \mu^+ \mu^-$ for various choices of $m_H$ and $m_{e_4}$ obtained by recasting the ATLAS search for $H \to \gamma \gamma$ with $N_{\ell} \ge 1$ ~\cite{Aad:2014lwa} from the 8 TeV full data.
}
\label{fig:dpmm8bound}
\end{center}
\end{figure}

The fiducial region adopted in ref.~\cite{Aad:2014lwa} to study the $\gamma \gamma \mu X$ final state is defined as follows. The diphoton event is selected when the invariant mass is in the range $105~{\rm GeV} \le m_{\gamma \gamma} < 160$ GeV and $p_T^\gamma > 0.35$ (0.25) of $m_{\gamma \gamma}$ for the leading (next-to-leading) photon. At least one isolated lepton with $p_T^\mu > 15$ GeV is requested. The majority of our signal events pass this inclusive lepton selection cut ($N_{\ell} \ge 1$) leading to large acceptance. 

From table 3 of ref.~\cite{Aad:2014lwa} the upper limit on the fiducial cross section is about 0.80 fb at 95\% C.L.. The limit on ${\rm BR}(H \to h \mu^+ \mu^-)$ is obtained from $A_{{\rm NP} \gamma \gamma} \times \sigma (p p \to H) \times {\rm BR}(H \to h \mu^+ \mu^-) \times {\rm BR}(h \to \gamma \gamma) < 0.80$ fb, and is presented in figure~\ref{fig:dpmm8bound} for various values of $m_H$ and $m_{e_4}$. We can see that the strength of this constraint is similar to that of the $b \bar b \mu^+ \mu^-$ search.

\subsection{Recast of the $\gamma \gamma \mu^+ \mu^-$ search}
\label{sec:dpmm8bis}
In ref.~\cite{Aad:2016sau} ATLAS presented a study of the $\gamma\gamma\mu^+\mu^-$ final state. The fiducial cuts adopted are $E_T^\gamma > 15$ GeV, $p_T^\mu > 25$ GeV, $m_{\mu\mu} > 40$ GeV and $\Delta R^{\gamma\gamma,\gamma \mu} > 0.4$. Muons and photons are required to be isolated from nearby hadronic activity within a cone of size $\Delta R = 0.4$. 

In order to place a constraint on our signal we consider only three bins with $m_{\gamma\gamma} \in [100,160]$ GeV (from the right panel of figure 4 of ref.~\cite{Aad:2016sau}). The observed number of events is 8 over a background of 5 (mainly from $pp\to Z(\mu^+\mu^-) \gamma\gamma$). The implied 95\% C.L. upper limit on a new physics signal is 9.6 events. 

\begin{figure}
\begin{center}
\includegraphics[width=.495\linewidth]{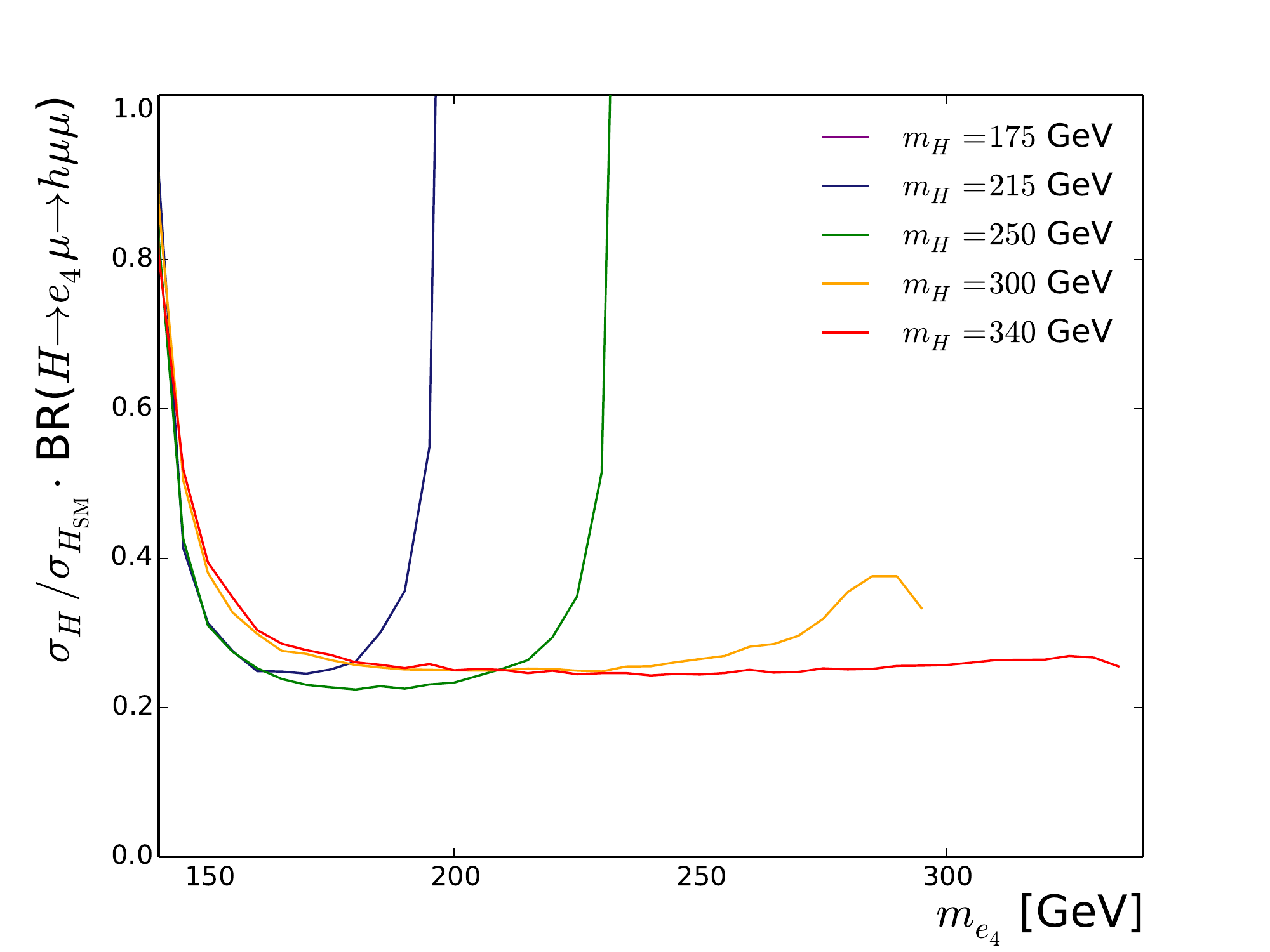}
\includegraphics[width=.495\linewidth]{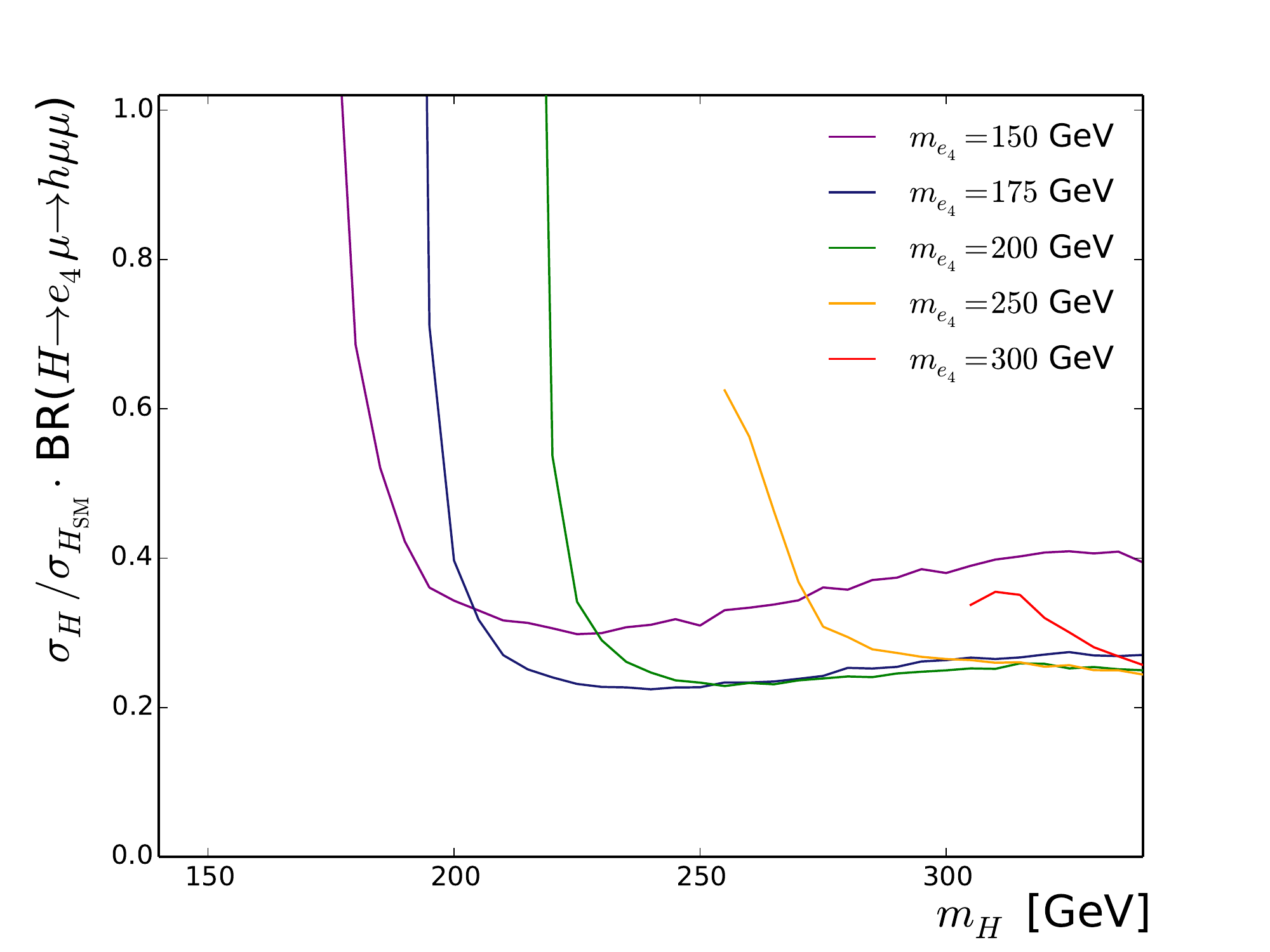}
\caption{The 95\% C.L. upper bounds on $pp \to H \to e_4^\pm \mu^\mp \to h \mu^+ \mu^-$ for various choices of $m_H$ and $m_{e_4}$ obtained by recasting the ATLAS search for $pp \to \gamma \gamma\mu^+\mu^-$ ~\cite{Aad:2016sau} from the 8 TeV full data.
}
\label{fig:dpmm8bisbound}
\end{center}
\end{figure}

Using a detector level efficiency of 37.7\% (as given in table 6 of ref.~\cite{Aad:2016sau}), we obtain the bounds shown in figure~\ref{fig:dpmm8bisbound}. These bounds are slightly worse than those from the $\gamma\gamma\mu X$ analysis in part because in that analysis the observed limit was slightly better than the expected one, while in the $\gamma\gamma\mu^+\mu^-$ analysis there was a small excess for $100 \; {\rm GeV} < m_{\gamma\gamma} < 160 \; {\rm GeV}$.

\section{Expected experimental sensitivities}
\label{sec:sensitivity}
\begin{figure}
\begin{center}
\includegraphics[width=.49\linewidth]{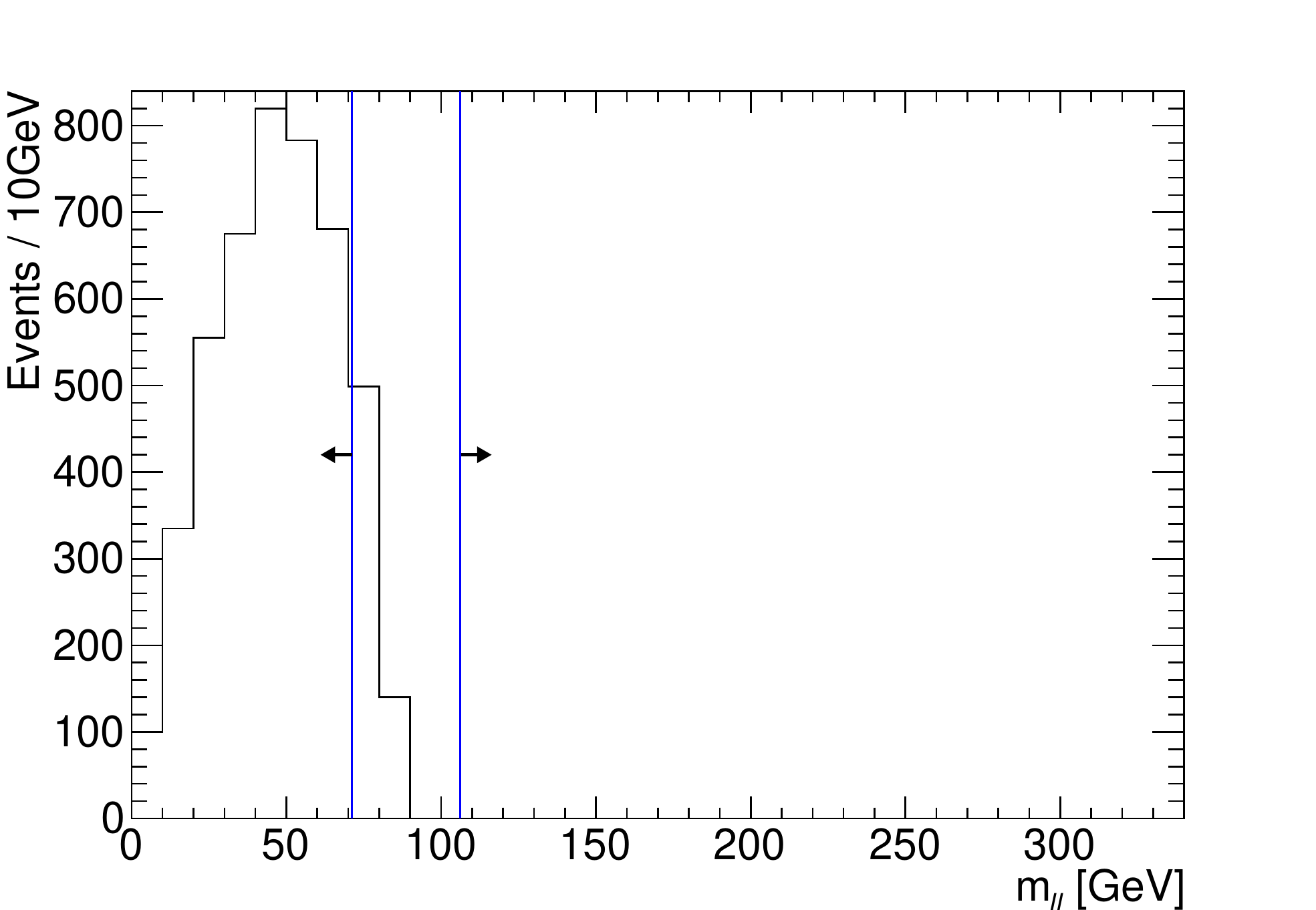}
\includegraphics[width=.49\linewidth]{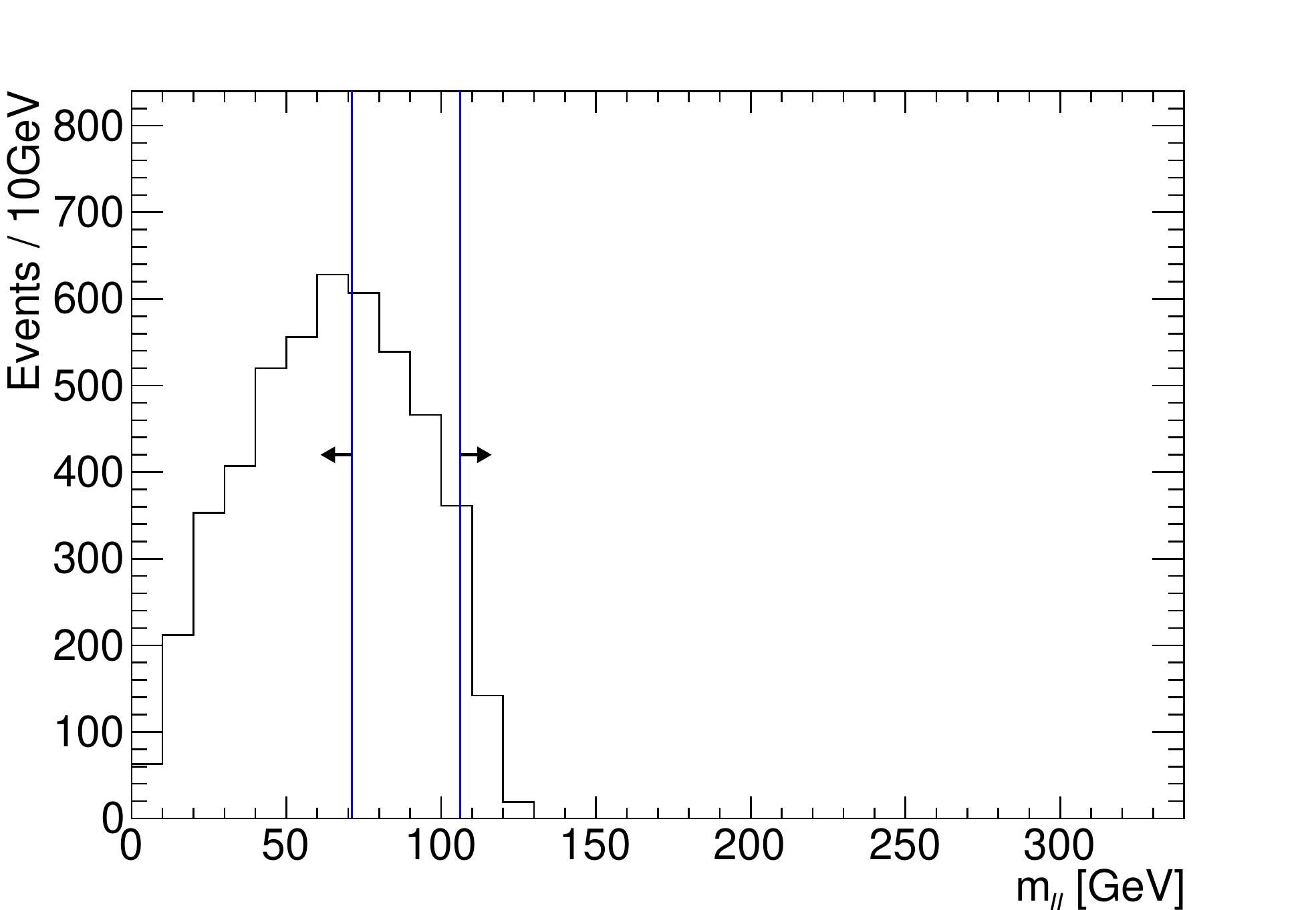}
\includegraphics[width=.49\linewidth]{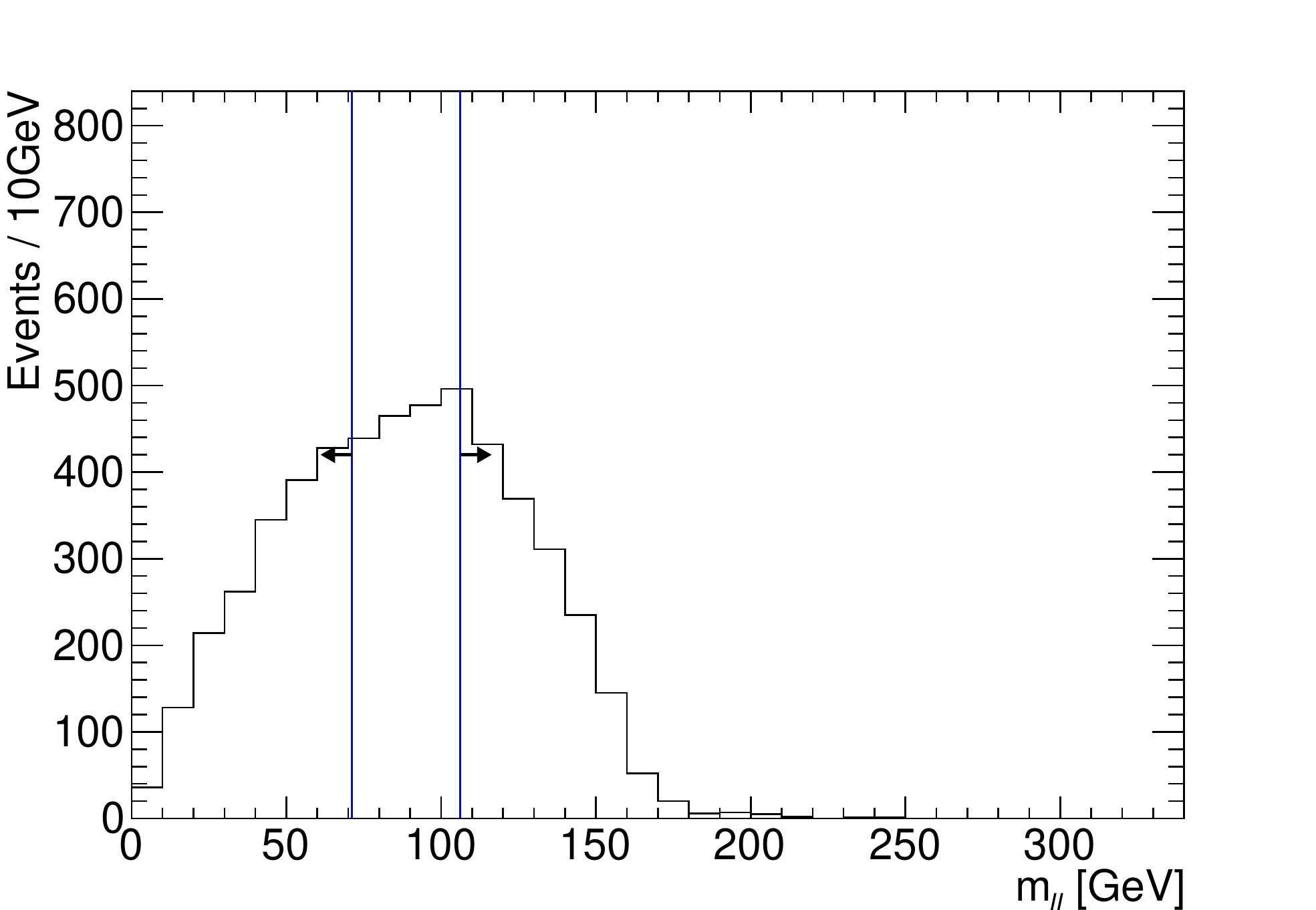}
\includegraphics[width=.49\linewidth]{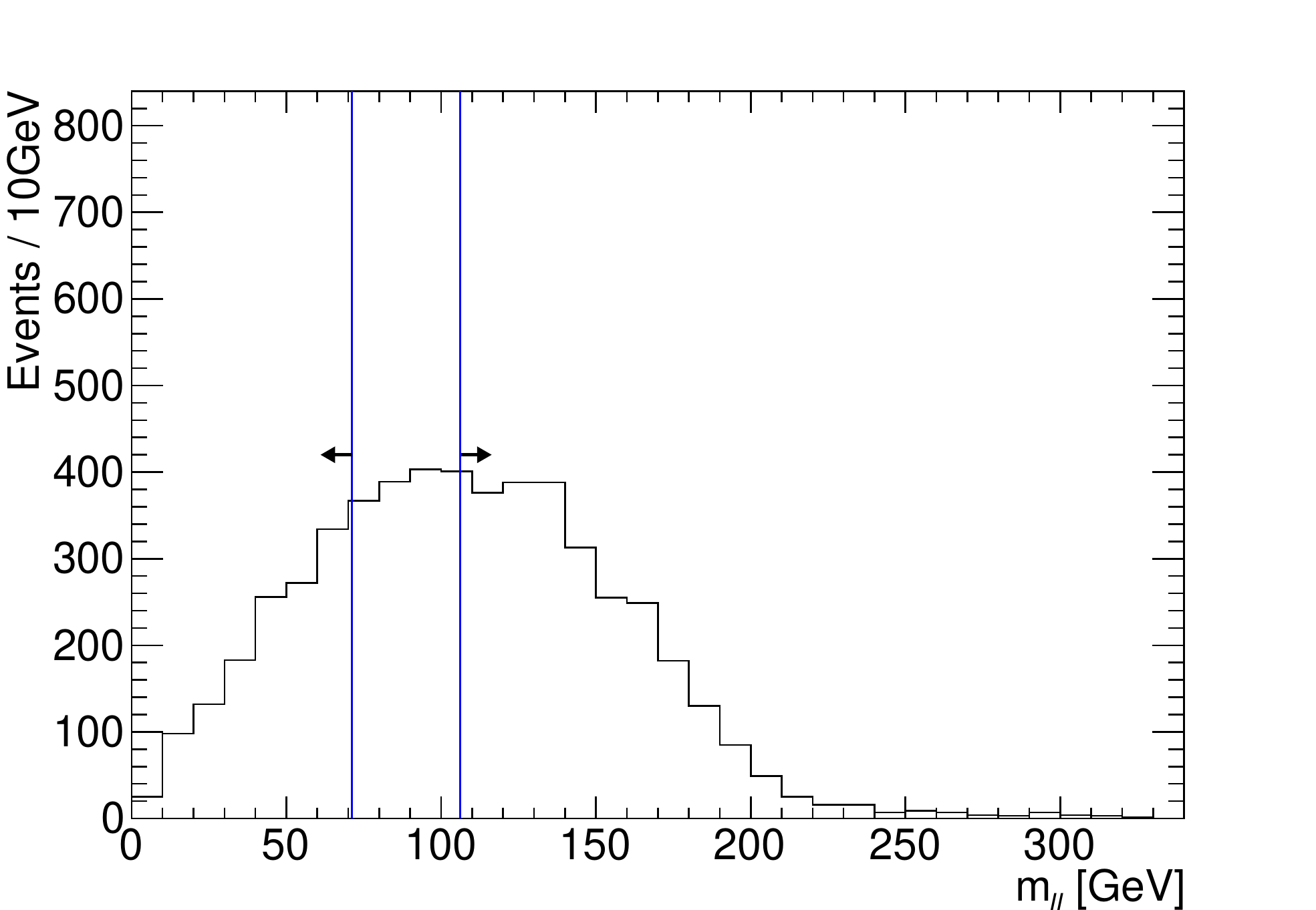}
\caption{$m_{\ell \ell}$ distributions for $m_H = 215,250,300,340$ GeV at fixed $m_{e_4} = 170$ GeV for the $b \bar b \mu^+ \mu^-$ channel. All the cuts described in section~\ref{sec:bbmm8} (with the exception of the $m_{\ell\ell}$ one) are imposed.}
\label{fig:mllex1}
\end{center}
\end{figure}

\begin{figure}
\begin{center}
\includegraphics[width=.49\linewidth]{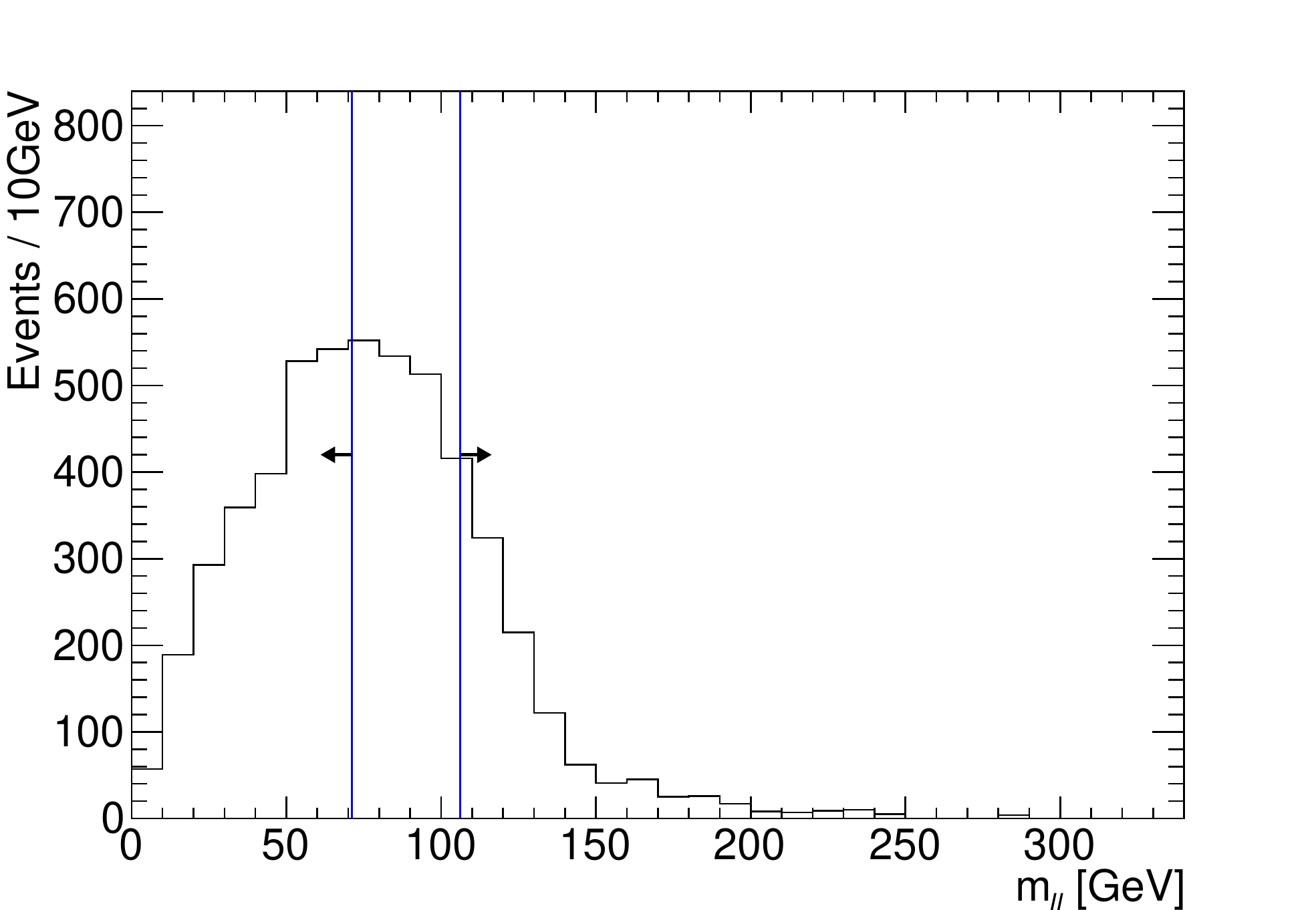}
\includegraphics[width=.49\linewidth]{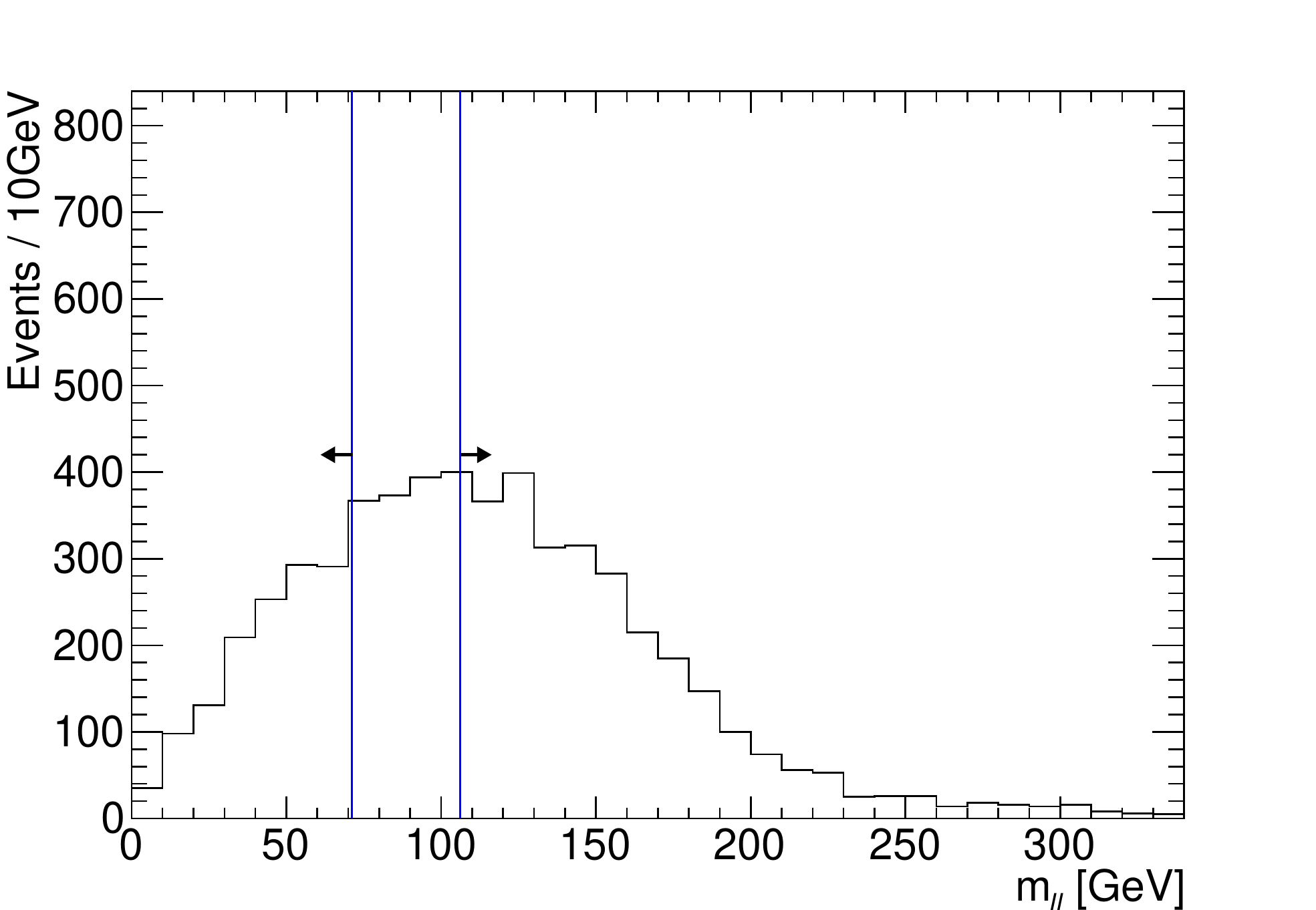}
\caption{$m_{\ell \ell}$ distributions for $m_H = 300,340$ GeV at fixed $m_{e_4} = 260$ GeV for the $b \bar b \mu^+ \mu^-$ channel. All the cuts described in section~\ref{sec:bbmm8} (with the exception of the $m_{\ell\ell}$ one) are imposed.}
\label{fig:mllex2}
\end{center}
\end{figure}

In this section we suggest new selection cuts to improve the sensitivity to our signal. First let us discuss the distribution of the invariant mass of the dilepton system $m_{\ell \ell}$. In our process the two oppositely charged muons are not produced from a $Z$ decay. The analytic formula for $m_{\ell \ell}$ is
\begin{align}
m_{\ell \ell} = \sqrt{\frac{(m_H^2 - m_{e_4}^2)(m_{e_4}^2 - m_h^2) (1 - \cos\theta)}{m_H^2 + m_{e_4}^2 + (m_H^2 - m_{e_4}^2) \cos\theta}}~,
\label{dilepton}
\end{align}
where $\theta$ is the angle between the two muons in the heavy Higgs rest frame. The maximum value of $m_{\ell \ell}$, obtained for $\cos\theta = -1$ and $m_{e_4} = \sqrt{m_H m_h}$, is $m_H - m_h$. The detailed distribution of $m_{\ell \ell}$ depends on the masses $m_H$ and $m_{e_4}$. Examples are shown in figure~\ref{fig:mllex1} for $m_H = 215,250,300,340$ GeV at fixed $m_{e_4} = 170$ GeV and figure~\ref{fig:mllex2} for $m_H = 300, 340$ GeV at fixed $m_{e_4} = 260$ GeV. 

As discussed in ref.~\cite{Dermisek:2015hue} a large part of our signal lies in the region $|m_{\ell \ell} - M_Z| > 15$ GeV allowing us to veto a major background process, $Z$ + (heavy flavored) jets with $Z \to \mu^+ \mu^-$. For this reason we propose to consider separately $m_{\ell \ell} > M_Z + 15$ GeV and $20 < m_{\ell \ell} < M_Z - 15$ GeV cuts (we added a lower limit $m_{\ell \ell} > 20$ GeV to suppress the background events with $\mu^+ \mu^-$ from $\gamma^\ast$). We call these cuts ``off-$Z$ below" and ``off-$Z$ above" cuts. In each panel of figures~\ref{fig:mllex1} and \ref{fig:mllex2} we show such regions with blue vertical lines and arrows. We see that for small $m_H - m_h$ and/or $m_H - m_{e_4}$ the ``above" cut is depleted of events.

For the $b \bar b \mu^+ \mu^-$ channel we keep the rest of the cuts in ref.~\cite{Aad:2015yza} other than $83 \; {\rm GeV} < m_{\ell \ell} < 99 \; {\rm GeV}$. Additionally we request that the invariant mass of all the final states $m_{\mu \mu b b}$ should be within 10\% of each $m_H$ hypothesis. Profile likelihood fits can be used once actual data are available.

For the $\gamma \gamma \mu^+ \mu^-$ channel we further impose a missing transverse energy cut $E_T^{\rm miss} < 60$ GeV to suppress the background from the top-quark decays. Moreover we request two leptons with $p_T > 15$ GeV.

\subsection{Sensitivity of $b \bar b \mu^+ \mu^-$}
\label{sec:newbb}
We begin by studying how the sensitivity of the existing 8 TeV 20 fb$^{-1}$ $b \bar b \mu^+ \mu^-$ search changes with the adoption of the new cuts we propose. This is controlled by the change in the expected number of background event that is obtained by computing the ratio of acceptances of new and original cuts:
\begin{align}
b = \frac{A_B^{\rm new}}{A_B^{\rm original}} \, b_0~,
\end{align}
where $b_0$ is the number of background events given in ref.~\cite{Aad:2015yza}, $A_B^{\rm new}$ and $A_B^{\rm original}$ are the MC level acceptances for the new and original cuts, respectively. The ratio of acceptances are obtained from the sample including $Z$ + $b$-jets, $t \bar t$, and Higgsstrahlung processes.\footnote{We do not include the subdominant background channels like single top quark and diboson ($VV$) production.} 

The 95\% CL$_{\rm s}$ median upper limits $N_s^{95}$ obtained from the number of expected background and the ratio of acceptances are shown in table~\ref{table:newbb} for our reference parameters $m_H = 215, 250, 300, 340$ GeV. We present separate results for the ``off-$Z$ below" and ''off -$Z$ above" cuts. Finally, the experimental sensitivities are obtained by inserting these limits in eq.~(\ref{eq:sensitivity}).

\begin{table}[htp]
\begin{center}
\begin{tabular}{|c||c|c|c|c|c|c|}
\hline
& \multicolumn{3}{|c|}{$N_s^{95}$ (``off-$Z$ below'')} & \multicolumn{3}{|c|}{$N_s^{95}$ (``off-$Z$ above'')} \\
\hline
$m_H$ [GeV] & 8 TeV        & 13 TeV        & 13 TeV      &  8 TeV       & 13 TeV        & 13 TeV \\
            & 20 fb$^{-1}$ & 100 fb$^{-1}$ & 3 ab$^{-1}$ & 20 fb$^{-1}$ & 100 fb$^{-1}$ & 3 ab$^{-1}$ \\ \hline
215 & 18 & 18 & 92  & 3  & 53 & 303 \\ \hline
250 & 21 & 64 & 342 & 13 & 95 & 516 \\ \hline
300 & 19 & 74 & 398 & 23 & 76 & 409 \\ \hline
340 & 7  & 74 & 398 & 15 & 42 & 225 \\
\hline
\end{tabular}
\end{center}
\caption{The expected upper limits for $b \bar b \mu^+ \mu^-$ searches.}
\label{table:newbb}
\end{table}%

To estimate the sensitivity at 13 TeV we start with considering the cuts used in the recent ATLAS analysis~\cite{ATLAS-CONF-2016-015} performed with 3.2 fb$^{-1}$ of integrated luminosity at 13 TeV. Since we are interested in $m_H < 340$ GeV for now, we consider the low $p_T^Z$ category. 

The basic cuts adopted in this search are the following. One of the two leptons must have $p_T > 25$ GeV with $|\eta| < 2.5$ and the invariant mass of the dilepton should be in the $70 \; {\rm GeV} < m_{\ell \ell} < 110 \; {\rm GeV}$ window. Events with two $b$-tagged jets are selected when one of them satisfies $p_T > 45$ GeV on top of their basic $b$-jet selection criteria. The invariant mass of the two $b$-tagged jets must be in the range $110 \; {\rm GeV} < m_{b \bar b} < 140 \; {\rm GeV}$. In order to suppress the $t \bar t$ background the missing transverse energy should be in the range $E_T^{\rm miss} / \sqrt{H_T} < 3.5 \, \sqrt{{\rm GeV}}$ where $H_T$ is the scalar sum of the $p_T$ of the leptons and $b$-tagged jets. To improve the resolution of the $b \bar b \ell^+ \ell^-$ resonance signal the four momentum of the $b \bar b$ system is rescaled by $m_h / m_{b \bar b}$ where $m_h = 125$ GeV as in the 8 TeV search~\cite{Aad:2015yza}. Because the main goal of the search in ref.~\cite{ATLAS-CONF-2016-015} is finding the resonant signal $A \to h Z$, the four momentum of the dimuon system is rescaled by $M_Z / m_{\mu \mu}$ with $M_Z = 91.2$ GeV: this requirement strongly suppresses the acceptance of our signal implying the absence of any constraint.

Our proposed cuts involve adding the ``off-$Z$  above" and ``off-$Z$ below" cuts described above, removing the rescaling of the four momentum of the dimuon system and including the invariant mass cut $|m_{\mu \mu b b} - m_H| < 0.1\,m_H$. The number of expected background events with integrated luminosities of 100 fb$^{-1}$ and 3 ab$^{-1}$ are calculated analogously to the 8 TeV case and the corresponding $N_s^{95}$ are summarized in table~\ref{table:newbb}.

\subsection{Sensitivity of $\gamma \gamma \mu^+ \mu^-$}
\label{sec:newdp}

The cuts that we suggest are those considered in ref.~\cite{Aad:2014lwa} (and described in section~\ref{sec:dpmm8}) with the inclusion of the ``off-$Z$ below''/``off-$Z$ above" cuts, a missing transverse energy cut $E_T^{\rm miss} < 60$ GeV (to suppress the $ht\bar t$ final state) and the requirement of a second isolated muon with $p_T > 15$ GeV. Additionally one could add a veto on high $p_T$ $b$-jets (for an additional suppression of the $ht\bar t$ background) and a cut on the invariant mass of the $\gamma\gamma\mu^+\mu^-$ system. The latter, in particular, could turn useful if non-irreducible sources of background turns out to be larger than expected. 

The background to the $\gamma \gamma \mu^+ \mu^-$ channel has been studied in detail in ref.~\cite{Aad:2016sau} ($E_\gamma > 15$ GeV and $p_T^\mu > 25$ GeV) and it is found to be dominated by $pp\to Z (\mu^+\mu^-)\gamma\gamma$ and $pp\to Z+\gamma j, j\gamma, jj$ with one or two jets misidentified as isolated photons. These backgrounds are also found to decrease steeply with the transverse energy of the photon. The $E_T^\gamma$ cuts that we suggest are much stronger (the hardest photon has $E_T^\gamma > 37-56$ GeV depending on the diphoton invariant mass) than those considered in ref.~\cite{Aad:2016sau} and make this background completely negligible (also taking into account the further reduction due to the off-$Z$ cuts). Two more sources of background (that are not suppressed by a stronger $E_T^\gamma$ cut) are $p p \to h Z \to \gamma \gamma \mu^+ \mu^-$ and $p p \to h t \bar t \to b \bar b \mu^+ \mu^- \gamma \gamma \nu_\mu \bar \nu_\mu$ (Presently we do not require vetos on $b$-jets, hence any $\gamma\gamma\mu^+\mu^- X$ final state is a background). At 8 TeV the combined total cross section for these two processes is about 35 ab corresponding to 0.7 events with 20 fb$^{-1}$ before applying any selection cut; therefore, we set this background to zero and find $N_s^{95} = 3$. At 13 TeV the combined cross sections rises to 80 ab corresponding to 8 and 240 events with 100 fb$^{-1}$ and 3 ab$^{-1}$, respectively; in this case a discussion of fiducial acceptances and detector efficiencies is crucial to estimate the expected background.

Using these selection cuts we find that the fiducial acceptances for the ``off-$Z$ below'' and ``off-$Z$ above" cases are 1.4\% and 1.2\%, respectively. Assuming an overall detector efficiency of about 37.7\% (as suggested in ref.~\cite{Aad:2016sau}), we then find that the expected number of background events at 13 TeV with 100 fb$^{-1}$ and 3 ab$^{-1}$ are 0 and 1, respectively: the corresponding $N_s^{95}$ are 3 and  4 events.

\section{Constraints and future prospects in two Higgs doublet model}
\label{sec:results}
%
\begin{figure}
\begin{center}
\includegraphics[width=.495\linewidth]{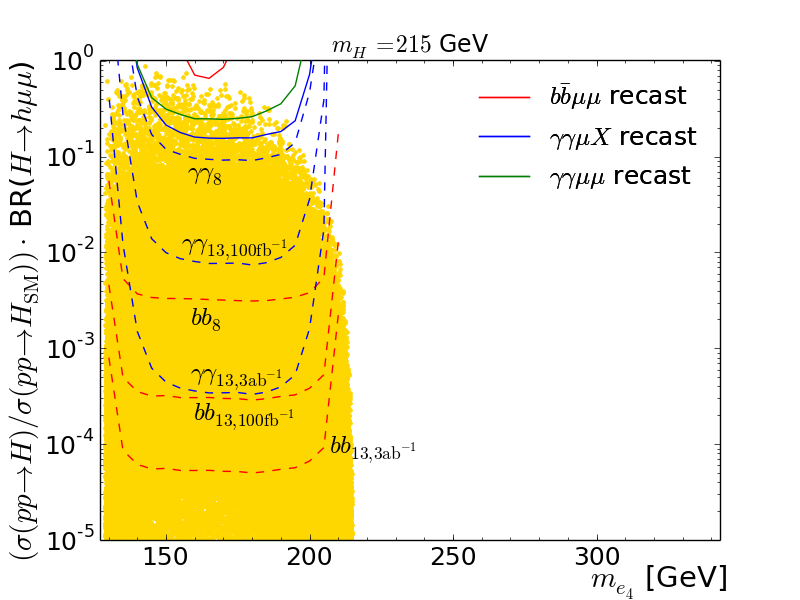}
\includegraphics[width=.495\linewidth]{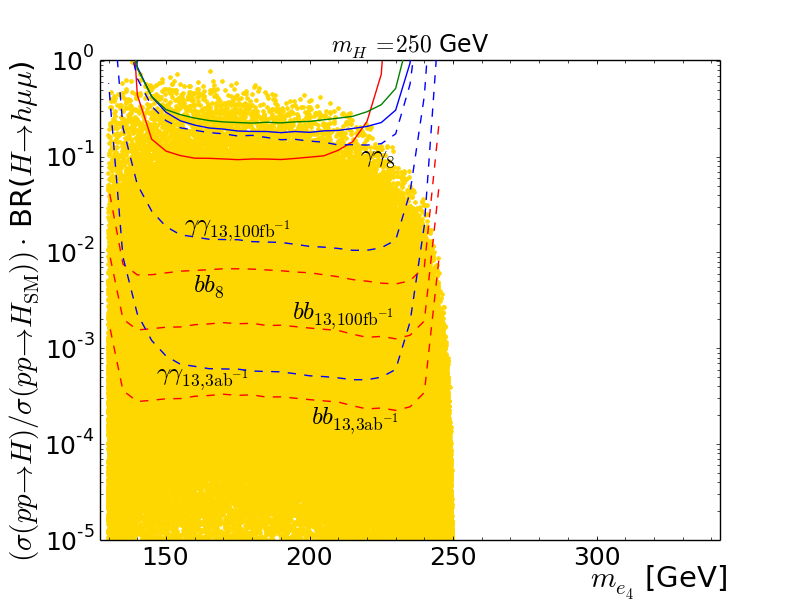}
\includegraphics[width=.495\linewidth]{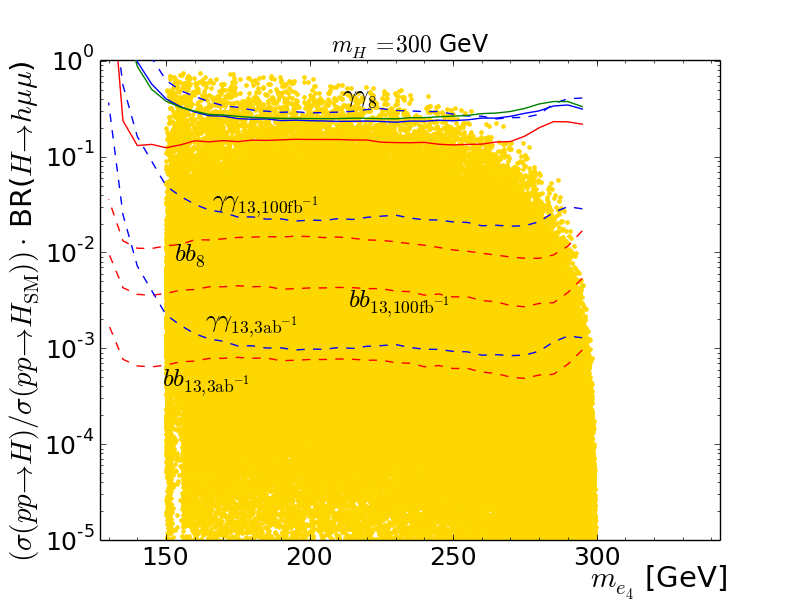}
\includegraphics[width=.495\linewidth]{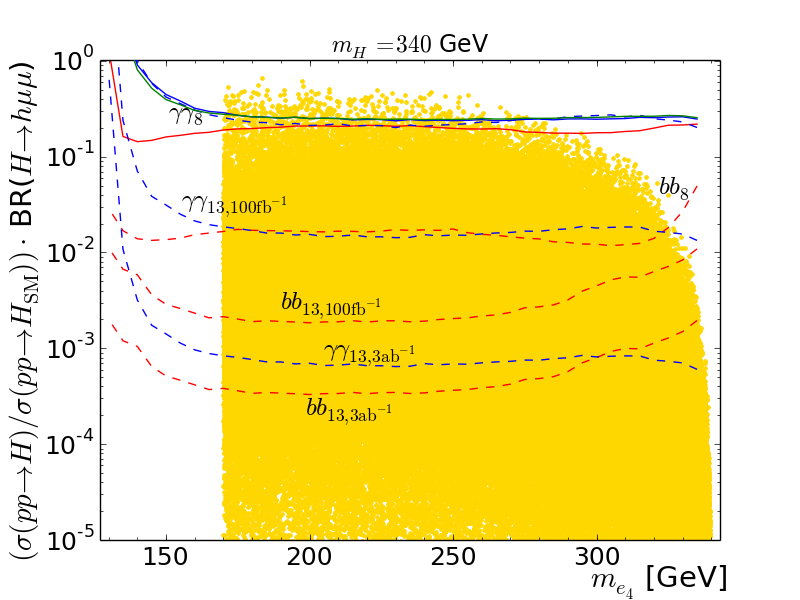}
\caption{The solid lines are the new constraints on the parameter space of the vectorlike lepton model discussed in ref.~\cite{Dermisek:2015oja} from $b\bar b \mu^+\mu^-$ (red), $\gamma\gamma\mu X$ (blue) and $\gamma\gamma\mu^+\mu^-$ (green) searches. The dashed lines are the expected constraints for the $b\bar b\mu^+\mu^-$ (red) and $\gamma\gamma\mu^+ \mu^-$ (blue) channels. We consider four heavy Higgs scenarios ($m_H = $ 215, 250, 300, 340 GeV) and three experimental setups ((8 TeV,20 fb$^{-1}$), (13 TeV,100 fb$^{-1}$) and (13 TeV,3 ab$^{-1}$)).
}
\label{fig:main}
\end{center}
\end{figure}

In this section we study the impact of the limits derived in previous sections (and indicate future prospects of existing and suggested searches) on the two Higgs doublet model type-II with vectorlike pairs of new leptons introduced in ref.~\cite{Dermisek:2015oja}. We assume that the new leptons mix only with one family of SM leptons and we consider the second family as an example. In figure~\ref{fig:main} we present the parameter space scan of this model in the plane spanned by $m_{e_4}$ and $\large ( \sigma(p p \to H) / \sigma(p p \to H_{\rm SM}) \large) \times {\rm BR}(H \to h \mu^+ \mu^-)$ for four different heavy Higgs masses ($m_H = $ 215, 250, 300, 340 GeV).\footnote{We calculate the cross sections assuming that $H$ is the heavy CP even Higgs. Note, however, that the limits are still model independent.} The charged sector Yukawa couplings are scanned in the range  [-0.5,\,0.5], as described in ref.~\cite{Dermisek:2015hue}. Each point satisfies precision EW data constraints related to the muon and muon neutrino: muon lifetime, $Z$-pole observables, the $W$ partial width and oblique observables. In addition, we impose constraints on pair production of vectorlike leptons obtained from searches for anomalous production of multilepton events~\cite{Dermisek:2014qca} and constraints from searches for heavy Higgs bosons in $H \to WW,\; \gamma \gamma$ discussed in ref.~\cite{Dermisek:2013cxa,Dermisek:2015vra, Dermisek:2015hue} and for the SM Higgs $h \to \gamma \gamma$ discussed in ref.~\cite{Dermisek:2015hue}.
 
The solid red, blue and green contours in figure~\ref{fig:main} are the new constraints obtained from recasting the existing $b\bar b\mu^+\mu^-$, $\gamma\gamma\mu X$ and $\gamma\gamma\mu^+ \mu^-$ searches. Note that the $\gamma\gamma\mu X$ and $\gamma\gamma\mu^+\mu^-$ constraints dominate at low $m_H$ because the $b \bar b \mu^+ \mu^-$ search looses sensitivity due to a strong cut on the transverse momentum of the hardest muon. Dashed contours indicate expected sensitivities using our proposed off-$Z$ cuts for three scenarios of LHC energies and integrated luminosities: (8 TeV,\; 20 fb$^{-1}$), (13 TeV,\; 100 fb$^{-1}$) and (13 TeV,\; 3 ab$^{-1}$). The contours shown correspond to the ``off-$Z$ below" cut for $m_H = $ 215 and 250 GeV and  ``off-$Z$ above" cut for $m_H = $ 340 GeV. For $m_H = $ 300 GeV both off-$Z$ cuts result in similar bounds. A direct inspection of figure~\ref{fig:main} shows that the analysis strategy we propose has the potential to improve the experimental sensitivity between one and two orders of magnitude depending on the heavy Higgs and vectorlike lepton masses.

From the sensitivities shown in figure~\ref{fig:main} we see that the impact of the off-$Z$ cuts is much more pronounced for the $b\bar b\mu^+\mu^-$ final state rather for the $\gamma\gamma\mu^+\mu^-$ one and the expected bounds converge at very high integrated luminosity. The reason is that the background to the existing $\gamma\gamma\mu^+\mu^-$ search is very small at all luminosities and, therefore, is not affected much by the additional off-$Z$ cuts; in the $b\bar b\mu^+\mu^-$ channel the background is large and is sizably reduced by the cuts we propose. At very large luminosity the expected number of background events increases much more for $b\bar b\mu^+\mu^-$ rather than $\gamma\gamma\mu^+\mu^-$ and the sensitivity of the two channels become comparable. At very high luminosities (beyond what is planned for the LHC) the di-photon channel would dominate.

Overall the potential for exclusion (discovery) of new physics in these channels in the next few years seems very strong: sensitivity to branching ratios of order $\mathcal{O}(10^{-4} - 10^{-3})$ is within reach and, correspondently, a very large part of this model parameter space will be tested. 

We should note that, in ref.~\cite{Aad:2014xzb}, ATLAS presented a search for $b\bar b \mu^+ \mu^-$ that makes use of multivariate techniques to massively reduce the irreducible background. While we were not able to use this analysis to place constraints on our model, we expect that a dedicated experimental study of the signal we propose using a similar approach has the potential to improve significantly the bounds we presented. The sensitivity could be additionally increased by looking for the $e_4 \to h\mu \to (b\bar b,\gamma\gamma)\mu$ resonance.

Finally let us briefly discuss the decay $H\to h\mu^+\mu^-$ with the SM Higgs decaying into the other possible channels we mention in table~\ref{table:channel}. The $h \to Z Z^\ast$ decay yields a $4\ell \mu^+ \mu^-$ final state that has negligible SM background; nevertheless the small branching ratio makes this channel less sensitive than the $\gamma\gamma \mu^+\mu^-$ one. On the other hand, the sizable $h\to\tau^+\tau^-$ branching ratio (about 6.3\%) makes the $\tau^+\tau^-\mu^+\mu^-$ final state competitive with the $b\bar b\mu^+\mu^-$ one; a detailed study of this final state from $p p \to A \to h Z$ has been performed by both ATLAS~\cite{Aad:2015wra} and CMS~\cite{Khachatryan:2015tha}. The $h\to W W^\ast$ mode yields the $2\ell2\mu 2\nu$ final state and is expected to yield sensitivities even higher than the $\gamma\gamma\mu^+\mu^-$ channel (both have negligible background and the former has a larger branching ratio). Finally the $h\to \mu^+\mu^-$ decay yields a $4\mu$ final state with a rate that depends strongly on the model Yukawa couplings (see the discussion in refs.~\cite{Dermisek:2013gta, Dermisek:2014cia}).

\section{Heavy Higgs above the top threshold}
\label{sec:heavyH}
%

\begin{figure}
\begin{center}
\includegraphics[width=.7\linewidth]{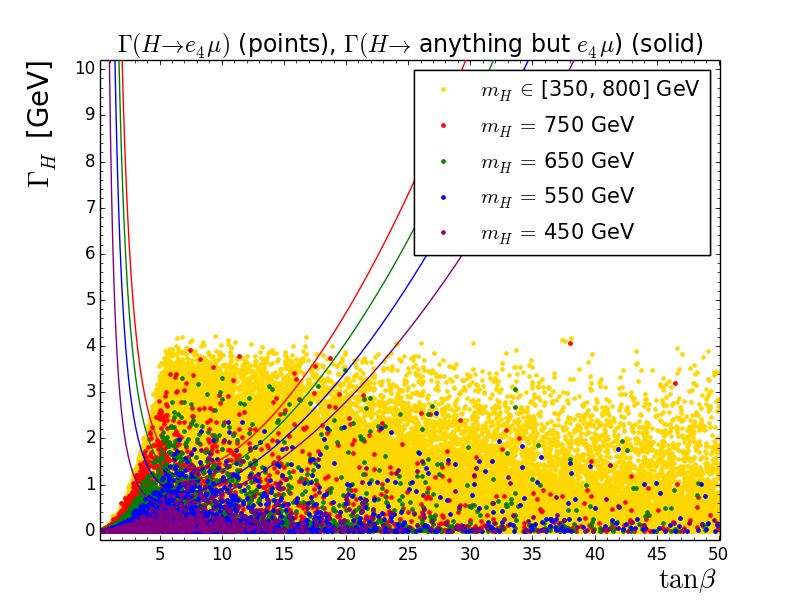}
\caption{
Heavy Higgs partial widths $\Gamma(H \to e_4^\pm \mu^\mp)$  for $m_H \in [350, 800]$ GeV are shown with yellow points. The widths for the reference Higgs masses $m_H =$ 450, 550, 650, and 750 GeV (purple, blue, green, and red colors) can be compared with $\Gamma(H \to$ anything but $e_4^\pm \mu^\mp)$ (solid lines).
}
\label{fig:heavyH1}
\end{center}
\end{figure}

\begin{figure}
\begin{center}
\includegraphics[width=.7\linewidth]{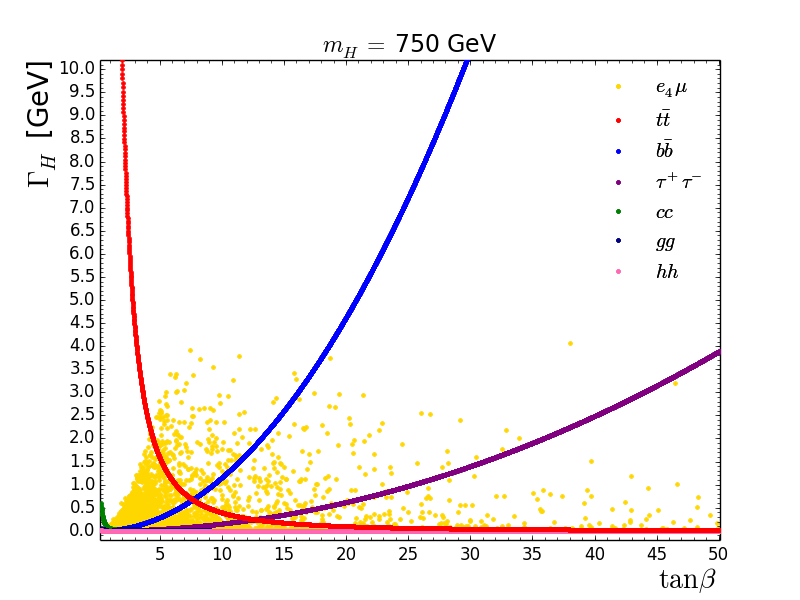}
\caption{
Heavy Higgs partial widths $\Gamma(H \to e_4^\pm \mu^\mp)$ for $m_H = 750$ GeV are shown with the yellow scattered points which correspond to the red points in figure~\ref{fig:heavyH1}. Partial widths of $H$ to various SM particles are shown with different colored lines.
}
\label{fig:heavyH2}
\end{center}
\end{figure}

In this section we discuss the constraints and prospects for $m_H \gtrsim 2 m_t$ where the $H\to t\bar t$ contribution to the heavy Higgs decay width reduces its branching ratio into vectorlike leptons. In this mass range, the heavy Higgs width into SM fermions is dominated by the $t\bar t$ channel at moderate $\tan\beta < 7$ and by the $b\bar b$ at larger $\tan\beta$. 

From the analysis presented in ref.~\cite{Dermisek:2015hue} (see bottom-left panel of figure~3 of that paper) it is clear that the $H\to e_4^\pm \mu^\mp$ branching ratio can easily be dominant for all values of $\tan\beta \lesssim 20$ and $m_H < 2 m_t$. This implies immediately that we expect BR($H\to e_4^\pm \mu^\mp$) to be sizable for Higgs masses above the $t\bar t$ threshold at large $\tan\beta$ (where the $H\to t\bar t$ partial width is suppressed with respect to the $H\to b\bar b$ one). For $\tan\beta < 7$ the $H\to t\bar t$ partial width becomes dominant and we need a detailed numerical calculation in order to assess the size of the $H\to e_4^\pm \mu^\mp$ branching ratio.

In order to check whether large BR($H\to e_4^\pm \mu^\mp$) are allowed, we rescan the parameters for $m_H$ above the $t \bar t$ threshold up to 800 GeV and allow only parameter space points that satisfy all the constraints discussed in ref.~\cite{Dermisek:2015hue}: electroweak precision data, anomalous multilepton production with missing $E_T$, SM Higgs data for $h \to \gamma \gamma$, and heavy Higgs searches in the $\gamma \gamma$ and $WW$ channels. As in the previous case the charged sector Yukawa couplings are scanned in the range  [-0.5,\,0.5].  

In figure~\ref{fig:heavyH1} we show the resulting heavy Higgs partial widths (calculated assuming $H$ is the heavy CP even Higgs) as a function of $\tan\beta$. The widths $\Gamma(H \to e_4^\pm \mu^\mp)$ for $m_H \in [350, 800]$ GeV are shown with the yellow scattered points. For comparisons of these widths with those for $H \to$ SM particles we consider four different representative $H$ masses $m_H = 450, 550, 650, 750$ (purple, blue, green, and red colors) out of the yellow points. For these reference $H$ masses the widths $\Gamma(H\to \text{anything but}\; e_4^\pm \mu^\mp )$ are shown with the solid lines; they are dominated by $H\to t\bar t$ for $\tan\beta \lesssim 7$ and by $H \to b \bar b$ for $\tan\beta \gtrsim 7$ which can be directly read from figure~\ref{fig:heavyH2} for a fixed $m_H$ = 750 GeV. From  inspection of the figure we see that the  $H\to e_4^\pm \mu^\mp$ decay mode can be dominant for $4 \lesssim \tan\beta \lesssim 17$. However, this region depends on the allowed range of  Yukawa couplings. Increasing the  range to [-1,\,1], the $H\to e_4^\pm \mu^\mp$ can  dominate for $4 \lesssim \tan\beta \lesssim 32$.

Note that searches for heavy CP even neutral Higgs are extremely challenging because, as explained in ref.~\cite{Jung:2015gta}, the $gg \to H\to t\bar t$ resonant peak can be destroyed by interference with the SM background (especially for $400 \; {\rm GeV} \lesssim m_H \lesssim \; {\rm 900}$ GeV and $\tan\beta < 15$ in the aligned two Higgs doublet model type-II). For the CP odd Higgs this effect leads to more dip-like signals in a large range of parameters but it is still hard to observe for $m_H<600$ GeV and $\tan\beta \lesssim 5$. If the heavy Higgs couples to vectorlike leptons, as in the models we consider, the $H\to e_4^\pm \mu^\mp$ channel offers a new and very promising avenue to discovery.

\begin{figure}
\begin{center}
\includegraphics[width=.7\linewidth]{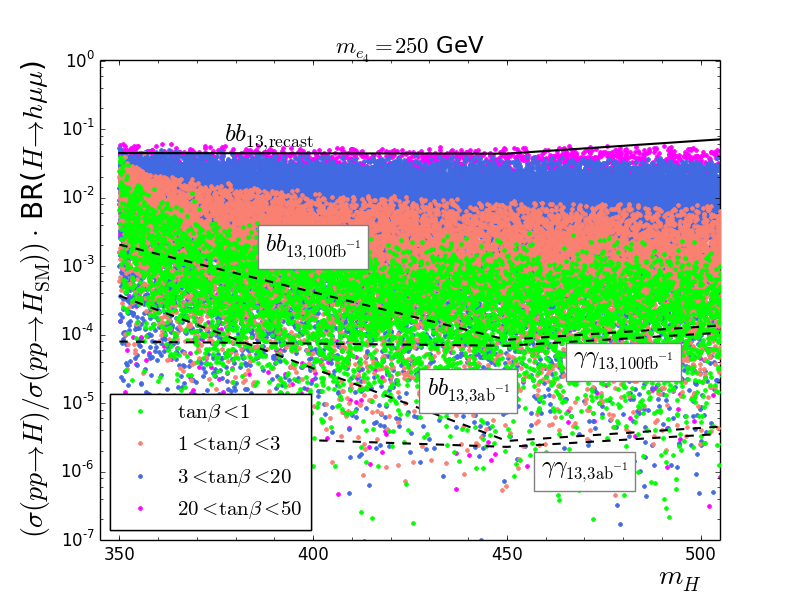}
\caption{
Parameter space satisfying all the constraints discussed in ref.~\cite{Dermisek:2015hue} for $m_{e_4} = 250$ GeV. We show estimates of the bound (black solid line) recasted from the 13 TeV $A \to hZ$ resonance search~\cite{ATLAS-CONF-2016-015} and future experimental sensitivities (black dashed lines) for integrated luminosities $\mathcal L = 300$ fb$^{-1}$ and 3 ab$^{-1}$ at 13 TeV.
}
\label{fig:heavyH3}
\end{center}
\end{figure}

In figure~\ref{fig:heavyH3} we show the allowed parameter space in the $m_H$ and $(\sigma_H / \sigma_{H_{\rm SM}}) \times {\rm BR}(H \to h \mu^+ \mu^-)$ plane. For simplicity we do not vary the vectorlike lepton mass and set it to $m_{e_4} = 250$ GeV; moreover we consider only the region $m_H < 2 m_{e_4}$ to kinematically forbid the $H\to e_4 e_4$ channel. Green, red, blue and magenta point correspond to $\tan\beta < 1$, $1 < \tan\beta < 3$, $3 < \tan\beta < 20$, and $20 < \tan\beta < 50$ respectively. From figure~\ref{fig:heavyH2} we see that our BR($H \to e_4^\pm \mu^\mp$) can be larger than 0.25 for $3 < \tan\beta < 20$. For larger $\tan\beta > 20$ the heavy Higgs production cross section is enhanced compared to $\sigma(p p \to H_{\rm SM})$ so the values of $(\sigma_H / \sigma_{H_{\rm SM}}) \times {\rm BR}(H \to h \mu^+ \mu^-)$ are as large as those for $3 < \tan\beta < 20$. The solid black contour is the recasted constraint from the 13 TeV $A \to hZ$ resonance search~\cite{ATLAS-CONF-2016-015}. The expected sensitivities of future $b\bar b \mu^+\mu^-$ and $\gamma\gamma\mu^+\mu^-$ studies are displayed as dashed lines. 

We conclude that recasted searches barely touch the allowed parameter space around $(\sigma_H / \sigma_{H_{\rm SM}}) \times {\rm BR}(H \to h \mu^+ \mu^-) \sim 0.05$. However, future searches employing the off-$Z$ cuts that we propose have the potential to constrain this quantity at $10^{-5}$ level.

\section{Conclusions}
\label{sec:conclusions}
In this paper we discuss the Higgs cascade decay $pp \to H \to e_4^\pm \mu^\mp \to h \mu^+ \mu^-$ that appears in models with extra vectorlike leptons and an extended Higgs sector. Among the various decay channels of the SM Higgs $h$ we considered the $b \bar b$ and $\gamma \gamma$ ones, which yield $b \bar b \mu^+ \mu^-$ and $\gamma \gamma \mu^+ \mu^-$ final states. These are two representative channels with sizable and negligible background, respectively. We were able to recast existing $pp \to A \to hZ \to b \bar b \ell^+ \ell^-$, $pp\to h\ell X \to \gamma \gamma \ell X$ and $pp \to Z \gamma \gamma \to \ell^+ \ell^- \gamma \gamma$ searches into constraints on the two modes we consider. We also presented the expected sensitivities of dedicated searches in the full 8 and 13 TeV data sets.  

A unique feature of cascade decay we consider is that the two leptons do not reconstruct a $Z$ boson, while the $h\mu$ and $h\mu\mu$ invariant masses peak at $m_{e_4}$ and $m_H$, respectively. Therefore, we suggest to employ two off-$Z$ cuts that focus on the region above and below the $Z$ resonance: $20 \; {\rm GeV} < m_{\ell \ell} < M_Z - 15$ GeV and $m_{\ell \ell} > M_Z + 15$ GeV. In addition to these suggested cuts, the searches for two resonances corresponding to the $H$ and $e_4$ masses will lead further to higher sensitivities. We find that this analysis strategy has the potential to improve the experimental sensitivity between one and two orders of magnitude depending on the heavy Higgs and vectorlike lepton masses.

We discussed an explicit realization of a new physics model in which this cascade decay is allowed to proceed with sizable branching ratio. The model has been introduced in ref.~\cite{Dermisek:2015hue} and involves a new family of vectorlike leptons and an extra Higgs doublet. We found that a vast majority of this model parameter space that survives various indirect and direct constraints can be easily tested by searches for heavy Higgs cascade decays. 

One major result of our analysis is that the $b\bar b \mu^+\mu^-$ channel dominates the $\gamma\gamma\mu^+\mu^-$ in most of the parameter space up to an integrated luminosity of 3 ab$^{-1}$ at 13 TeV. We also briefly discussed other possible channels and found that the $\tau^+ \tau^- \mu^+ \mu^-$ and $2\ell 2\mu 2\nu$ have the potential to offer constraints comparable to those obtained from the $b\bar b \mu^+\mu^-$ and $\gamma\gamma\mu^+\mu^-$ modes.

Furthermore we  discuss the reach of our search strategy for a heavy Higgs with mass above the $t\bar t$ threshold. We find that the $H\to e_4^\pm \mu^\mp$ branching ratio can  dominate over both $H \to t \bar t$ and $H \to b \bar b$ for $4 \lesssim \tan\beta \lesssim 17$ ($4 \lesssim \tan\beta \lesssim 32$) when charged sector Yukawa couplings are allowed in  [-0.5,\,0.5] ([-1,\,1]). However, even in the range of parameters where our process has only a small branching ratio, it can be the most promising search channel since the usual search strategies for $H \to t \bar t$ suffer from interference effect with the SM background. Rough estimates of future experimental sensitivities are extremely promising.

\section*{Acknowledgements}
EL thanks Narei Lorenzo Martinez for discussions on the $\gamma\gamma\mu^+\mu^-$ channel. SS thanks Bogdan A. Dobrescu and Zhen Liu for useful discussions. SS also thanks Fermi National Accelerator Laboratory for hospitality and support during the completion of this work. The work of RD and EL was supported in part by the U.S. Department of Energy under grant number {DE}-SC0010120. RD is supported in part by the Ministry of Science, ICT and Planning (MSIP), South Korea, through the Brain Pool Program. SS is supported in part by BK21 plus program.


\end{document}